\documentclass[3p,review]{elsarticle}

\usepackage{lineno,hyperref}
\usepackage{times}  
\usepackage{helvet} 
\usepackage{courier}  
\usepackage{graphicx} 

\usepackage{graphics, epstopdf, epsfig}
\usepackage{amstext}
\usepackage{subcaption}
\usepackage{multirow}
\usepackage{amsmath}
\usepackage{amssymb}
\usepackage{multicol}
\usepackage{rotating}
\usepackage{booktabs}
\usepackage{algorithm}
\usepackage{algpseudocode}
\usepackage{fixltx2e}
\usepackage{xcolor}
\usepackage{xspace}

\newcommand{\ie}{\emph{i.e.,}\xspace}
\newcommand{\eg}{\emph{e.g.,}\xspace}

\journal{Journal of KBS}

\begin{document}

\begin{frontmatter}

\title{Hierarchical Text Interaction for Rating Prediction}


\author[address1]{Jiahui Wen}
\ead{wen\_jiahui@outlook.com}

\author[address2]{Jingwen Ma\corref{mycorrespondingauthor}}
\ead{majingwei0824@gmail.com}

\author[address3]{Honghui Tu}
\ead{tuhkjet@foxmail.com}

\author[address4]{Mingyang Zhong}
\ead{my.zhong@hotmail.com}

\author[address1]{Guangda Zhang}
\ead{zhanggd\_nudt@hotmail.com}

\author[address4]{Wei Yin}
\ead{yinweihappy168@yahoo.com}


\author[address1]{Jian Fang}
\ead{fangjian\_alpc@163.com}


\cortext[mycorrespondingauthor]{Corresponding author}


\address[address1]{National Innovative Institute of Defense Technology, Beijing, China}
\address[address2]{Shandong Normal University, Jinan, Shangdong}
\address[address3]{National University of Defense Technology, Changsha, China}
\address[address4]{The University of Queensland, Australia}

\begin{abstract}
Traditional recommender systems encounter several challenges such as data sparsity and unexplained recommendation. To address these challenges, many works propose to exploit semantic information from review data. However, these methods have two major limitations in terms of the way to model textual features and capture textual interaction. For textual modeling, they simply concatenate all the reviews of a user/item into a single review. However, feature extraction at word/phrase level can violate the meaning of the original reviews. As for textual interaction, they defer the interactions to the prediction layer, making them fail to capture complex correlations between users and items. To address those limitations, we propose a novel Hierarchical Text Interaction model(HTI) for rating prediction. In HTI, we propose to model low-level word semantics and high-level review representations hierarchically. The hierarchy allows us to exploit textual features at different granularities. To further capture complex user-item interactions, we propose to exploit semantic correlations between each user-item pair at different hierarchies. At word level, we propose an attention mechanism specialized to each user-item pair, and capture the important words for representing each review. At review level, we mutually propagate textual features between the user and item, and capture the informative reviews. The aggregated review representations are integrated into a collaborative filtering framework for rating prediction. Experiments on five real-world datasets demonstrate that HTI outperforms state-of-the-art models by a large margin. Further case studies provide a deep insight into HTI's ability to capture semantic correlations at different levels of granularities for rating prediction.

\end{abstract}


\end{frontmatter}

\section{Introduction}
With the prevalence of Internet, online services generate massive amounts of data on a daily basis, and users are facing the problem of information overload when finding their interested items (e.g. books, movies). To address this problem, various recommendation techniques are proposed to quickly find relevant information for the users \cite{Zhang2017A}. Collaborative filtering (CF) \cite{su2009survey} \cite{koren2008factorization} is one of the most widely employed techniques for recommendation. It learns latent factors of user and item based on interaction records such as ratings \cite{mnih2008probabilistic} \cite{lee2001algorithms}. However, CF-based methods are mainly based on interaction data, and the learned latent factors do not explain why an item is recommended \cite{He2015TriRank}. Therefore, those methods encounter some challenges, such as data sparseness and lack of interpretation.

To address those limitations, researchers resort to learning semantic features \cite{TMM2019_cite_leizhu} from the reviews. In most online services, users are allowed to write reviews to express their opinions towards the consumed items. The reviews contain plenty of information that reveals user interests and item features. Prior works leverage topic modeling techniques such as Latent Dirichlet Allocation (LDA) \cite{blei2003latent} to extract high-level features based on the reviews \cite{wang2011collaborative} \cite{mcauley2013hidden}. Those methods demonstrate improved performance over latent factor models. However, they represent the reviews as bags of words, and fail to capture textual features and local semantics \cite{TMM2017_cite_leizhu} expressed at the phrase and sentence level.

Recently, deep learning techniques (e.g. convolutional neural networks) have been widely applied in text-related tasks such as text classification \cite{mikolov2013distributed}. Existing works demonstrate the effectiveness of the deep learning techniques in extracting high-level abstract representations from texts for boosting specific tasks such as recommendation. In the light of this, many works \cite{kim2016convolutional} \cite{zheng2017joint} leverage neural networks to derive high-level textual factors over the reviews, and integrate those features into latent factor models \cite{rendle2010factorization} for rating prediction. Some research \cite{wu2019context} \cite{liu2019daml} propose to capture complex user-item interaction by modeling customized textual features given the specific users/items.  Although these research demonstrate better recommendation performance than topic-based models, they have several limitations \cite{TOIS2019_cite_leizhu}. First, most of the review-based models \cite{chen2018neural} \cite{seo2017interpretable} defer the user-item interactions till the prediction layer, resulting in static latent features of users and items. Therefore, in those models, users/items are always represented with the same latent factors regardless of the target items/users. They are incapable of capturing dynamic correlations between users and items, and fail to model the complex interactions between them. Second, other review-based models \cite{wu2019context} \cite{liu2019daml} simply concatenate all reviews of each user/item into a single review. The irrelevant words may introduce noises and undermine the recommendation performance. Furthermore, their feature interrelations are modeled at word or phrase level, which can twist the meaning of the original sentences \cite{ribeiro2016should}. For example, it is not uncommon for an overall positive review to have a few negative words. Similarly, positive words can occasionally appear in reviews that present overall negative sentiment.

To address those limitations, we propose an HTI model that exploits textual features at different levels of hierarchies for item recommendation. In HTI, feature extractions are achieved hierarchically at both word and review levels. One advantage of the hierarchical architecture is that we can model local informativeness and global overall semantics of the reviews. The local and global information can complement each other for preserving rich textual features at different granularities. In addition, to uncover complex user-item interrelations, we propose to model semantic correlations between the users and items at different levels. Specifically, at word level, we propose an attention mechanism to summarize the words, where the attention scores are parameterized by each user-item pair.  At review level, we propose an interactive network that propagates textual features between the user and item, and guide each other to select the relevant reviews for recommendation.

Compared with existing methods, we model local word informativeness and global review sentiment with a hierarchical architecture. The hierarchical review information allows us to sufficiently exploit textual features at different levels of granularities. Further, rather than representing users and items with static latent features, we dynamically identify informative textual features at both word and review levels for each specific user-item pair. We conduct experiments on HTI over five real-world datasets with reviews. The experimental results demonstrate the advantage of HTI over state-of-the-art baselines, and validate the rationale underlying the hierarchical architecture of the proposed model. In summary, the contributions of this work are listed as follows:
\begin{itemize}
\item We propose a review-based recommendation model (HTI for short)with hierarchical text modeling. The model preserves rich textual information at both word and review levels, allowing it to exploit textual features at different levels of granularity for accurate rating prediction. 
\item We propose to model texts interaction in the hierarchical architecture, enabling it to capture informative local words and global reviews, and extract interaction-based textual features for the specific task of recommendation. In this sense, we are the first to simultaneously model hierarchical and interactive textual features for recommendation.

\item We conduct experiments on five real-world datasets,  and demonstrate that HTI consistently outperforms the state-of-the-art baselines by a large margin. We also study the contribution of each component for the recommendation task. Finally, case studies provide a better understanding of HTI's effectiveness in modeling textual interaction at different levels.

\end{itemize}


\section{Related Work}
Our work is related to recommendation with reviews, attention-based recommendation and hierarchical representation learning. We briefly review the recent advances in these areas.
\subsection{Recommendation with Reviews}
Traditional collaborative filtering recommenders \cite{Wang2019Neural} \cite{Wang2019Neural} \cite{parvin2019a} \cite{zhang2019a} are mainly based on user-item interactions. Therefore, they have two major limitations, namely data sparsity and cold-start problem, as users usually give few ratings given the larger number of items \citep{DBLP:journals/tcyb/HanZCLL20}. To tackle these limitations, many researches resort to extra knowledge such as social networks \cite{chen2019efficient} \cite{fan2019graph} \cite{wu2019neural}, images \cite{zhang2019scalable} \cite{zhang2017joint}. Among those auxiliary data sources, review texts have received much research attention \cite{TMM_cite_leizhu2020}. Review texts are ubiquitous in recommender systems, and there are strong inherent correlations between the textual sentiment and user preferences. Early researches employ topic modeling such as LDA \cite{blei2003latent} to uncover latent topic factors, and combine the topical information with ratings to recommend items \cite{ling2014ratings} \cite{bao2014topicmf}. CDL \cite{wang2015collaborative} leverages SDAE \cite{vincent2010stacked} to extract deep features from texts, and uses Bayesian model \cite{koren2009matrix} to achieve the combination of ratings and reviews for rating prediction. Those methods demonstrate better performance than the baselines that are mainly based on user-item rating data. However, they regard the reviews as bags of words (BOW), hence they fail to capture word-order information and local semantics \cite{Jie2020Generalized}.

To approach this limitation, several studies \cite{zhang2016collaborative} leverage deep neural networks such as CNNs for extracting high-level text features. Neural networks\cite{TNNLS_cite_leizhu2020} are proved to be powerful in text-related tasks such as text classification \cite{kim2014convolutional} \cite{kalchbrenner2014convolutional}, as they are able to capture local semantics of different granularities when filter maps of different sizes are applied. Kim et al. \cite{kim2016convolutional} propose a neural network model, ConvMF, for recommendation. ConvMF utilizes CNNs to extract item latent features from item reviews, and integrate those features into a Probabilistic Matrix Factorization (PMF) framework for rating predictions. DeepCoNN \cite{zheng2017joint} uses two parallel neural networks to learn user and item latent features from their respective reviews. Then the Factorization Machines (FM) \cite{rendle2010factorization} model is used to model the high-order interactions between the latent features of users and items. D-attn \cite{seo2017interpretable} models latent features of users/items with dual local and global attention. The neural networks are applied to the user/item reviews separately, and the resultant factors are used for rating prediction in a similar way as matrix factorization. NARRE \cite{chen2018neural} explores the usefulness of the reviews, and represents latent factor of each user/item as a weighted sum over the review representations. Those deep models demonstrate improved performance over the above BOW-based models \cite{TNNLS2018_cite_leizhu}. However, they defer the interactions between users and items till the prediction layer, and each user/item always has the same latent features regardless of the target item/user. Therefore, they fail to model complex correlations between users and items. To deeply exploit the interrelations between latent features of users and items, recent works \cite{wu2019context} \cite{liu2019daml} consider early interactions between textual features before deriving latent features of users and items. Specifically, they first capture contextual features at each position of the texts with convolution operations. Instead of max pooling over the features, they model the mutual correlations between contextual features of users and items, and apply convolution with mean pooling operations to calculate the final latent features of users and items. Despite proven to be effective in rating predictions, those methods simply concatenate all review texts of a user into a single review, which can inevitably introduce noises and incur information loss, due to the various lengths and numbers of user/item reviews. Furthermore, in those works, semantic correlations are modeled at word/phrase level, which may violate the original meaning of the sentences \cite{ribeiro2016should}.

\subsection{Attention-based Recommendation}
Attention Mechanism \cite{bahdanau2014neural} has wide applications \cite{yang2016hierarchical} in neural networks \cite{li2020discriminative} \cite{zhang2019binary} community. The motivation is to place higher weights to parts of information that is strongly related to the specific applications, and ignore the others that may negatively affect the model performance. In review-based area, as not all words have equal contribution, we need to pay attention to important words that have strong indications of user preferences and item properties. Seo et al. \cite{seo2017interpretable} employ a combination of local and global attention mechanism to identify important words. The local attention selects informative words from a local window, while the global attention filters out noisy words and helps to model overall semantics of the reviews. In \cite{chen2018neural}, the authors propose to learn usefulness of the reviews with an attention mechanism. For each item/user, the attention score of each review is proportional to the relevance between the review and user/item associated with the review. Lu et al. \cite{lu2018coevolutionary} utilize a combination of PMF and an attention-based GRU network to collaboratively learn latent factors of users and items from ratings and reviews. Wang et al. \cite{wang2017dynamic} propose an attention mechanism to aggregate multiple predictive models for article recommendation. CARL \cite{wu2019context} utilizes an attention mechanism to mutually model pair-based semantic relevance between textual features of users and items. DAML\cite{liu2019daml} integrates local and mutual attention of convolution network to jointly extract features from texts.

Compared with the above methods, we propose a hierarchical attention mechanism to preserve informative textual features at both word and review levels. At word level, the attentions are specialized by each user-item pair, hence we can identify useful words guided by user preferences and item characteristics. At review level, we model the semantic alignments \cite{TNNLS2019_cite_leizhu} between user and item reviews through an interactive network.

\subsection{Hierarchical Representation Learning}
Hierarchical representation learning \cite{ma2019hierarchical} is to learn latent features at a hierarchical manner, and capture user interests or item characteristics at different granularities for fine-grained recommendation. Chen et al. \cite{chen2017attentive} propose a hierarchical model that consists of component-level and item-level representation learnings. At component level, an attention mechanism is proposed to select informative components for representing items. At item level, another attention module is used to capture important items for comprising user interests. Wu et al. \cite{wu2019hierarchical} model user latent preferences for image recommendation by identifying three aspects (i.e. upload coherence, social influence and owner admiration). A hierarchical attention network is designed to capture both element and aspect level features. In \cite{cao2018attentive}, the authors approach group level and user level recommendation. They propose an attention mechanism to select expert users for representing the groups. Specifically, each group places high weights to a specific set of users who are confident with the target item. With the group information, the user preferences are modeled at a more general level to alleviate data sparsity \cite{Wang2016Member}. DBRec \cite{ma2019dbrec} iteratively performs collaborative filtering and latent group discovery. These two modules can complement each other and learn better user/item representations. In addition to the user-item interaction, it interacts user/item with item/user group to capture user preferences and item properties at different levels of granularities. HRM \cite{wang2015learning} hierarchically captures both sequential behavior and users' general tastes for next item recommendation.

The proposed work is related to hierarchical representation learning, as we model textual features both at word and review levels. However, the aforementioned methods linearly fuse the latent features at each level. On the contrary, we propose to model complex user-item interrelations at different hierarchies. 

\section{Problem Formalization}
In a recommendation problem, we denote $U=\{u_1,\cdots,u_M\}$ as a set of $M$ users, and $V=\{v_1,\cdots,v_N\}$ a set of $N$ items. The ratings of the users over the items are represented with a rating matrix $\mathbf{R}\in \mathbb{R}^{M\times N}$, where each entry $r_{ij}$ in the matrix denotes user $u_i$'s preference over item $v_j$. As we focus on explicit recommendation in this work, the ratings take explicit real values in a specific range (i.e. [1,5]). For each user-item pair  $(u_i,v_j)$, we denote $D_{u_i}=\{d_{i,1},\cdots,d_{i,m}\}$ as the repository of reviews written by the user, and $D_{v_j}=\{d_{j,1},\cdots,d_{j,n}\}$ as the set of reviews written about the item, where $m$ and $n$ denote the number of reviews of $u_i$ and $v_j$ respectively.

Given a user-item pair and their respective historical reviews, the recommendation problem is to exploit the textual information in the reviews and produce the estimated rating $\hat{r}_{ij}$ of $u_i$ over $v_j$. In the reminder of this paper, we denote matrices with upper case bold letters, vectors with lower case bold case letters without any specification, and scalars with non-bold letters. The major symbols used in the paper are summarized in Table \ref{tb:symbol}, other intermediate variables are detailed in the corresponding sections.

\begin{table}\footnotesize
\centering
\caption{Symbols and Descriptions}
\begin{tabular}{lp{5cm}}
\toprule
Symbol & Description \\
\hline
$u_i, v_j$ & notations of a user and a item respectively\\
$U=\{u_1,...,u_M\}$ & a set of $M$ users\\
$V=\{v_i,...,v_N\}$ & a set of $N$ users\\
$\mathbf{u}_i,\mathbf{v}_j$ & latent factors of $u_i$ and $v_j$ respectively\\
$\mathbf{R}$ & rating matrix\\
$r_{ij}, \hat{r}_{ij}$ & ground truth and predictive rating score of $u_i$ over $v_j$ respectively \\
$D_{u_i}=\{d_{i,k},...,d_{i,m}\}$ & a set of $u_i$'s $m$ reviews\\
$D_{v_j}=\{d_{j,1},...,d_{j,n}\}$ & a set of $v_j$'s $n$ reviews\\
$d=\{w_i\}_{i=1}^p$ & an example review with $p$ words\\
$\{\mathbf{w}_i\}_{i=1}^p$ & embeddings of words $\{w_i\}_{i=1}^p$\\
$\mathbf{d}$ & review representation of an example document $d$\\
$\{\mathbf{d}_{i,k}\}_{k=1}^m$ & the set of representations of $u_i$'s reviews\\
$\{\mathbf{d}_{j,t}\}_{t=1}^n$ & the set of representations of $v_j$'s reviews\\
$\mathbf{p}_i$ & initial representation of $u_i$'s reviews\\
$\mathbf{q}_j$ & initial representation of $v_j$'s reviews\\
$\mathbf{s}_i$ & intermediate representation of $u_i$'s reviews\\
$\mathbf{t}_j$ & intermediate representation of $v_j$'s reviews\\
$\mathbf{d}_i$ & aggregated representation of $u_i$'s reviews\\
$\mathbf{d}_j$ & aggregated representation of $v_j$'s reviews\\

$\phi_l$ & $l$-th layer neural network at the prediction module \\
$\Theta$ & all model parameters\\
$\lambda$ & regularization parameter\\
$\mathcal{D}$ & training set\\
$\eta$ & learning rate\\
\bottomrule
\end{tabular}
\label{tb:symbol}
\end{table}

\section{The Proposed Model}
\subsection{Overview}
%

\begin{figure*}
  \centering
  \subcaptionbox{(a)Overall architecture of the proposed model\label{subfig:model}}[1.\linewidth]
  {%
    \includegraphics[width=.6\linewidth]{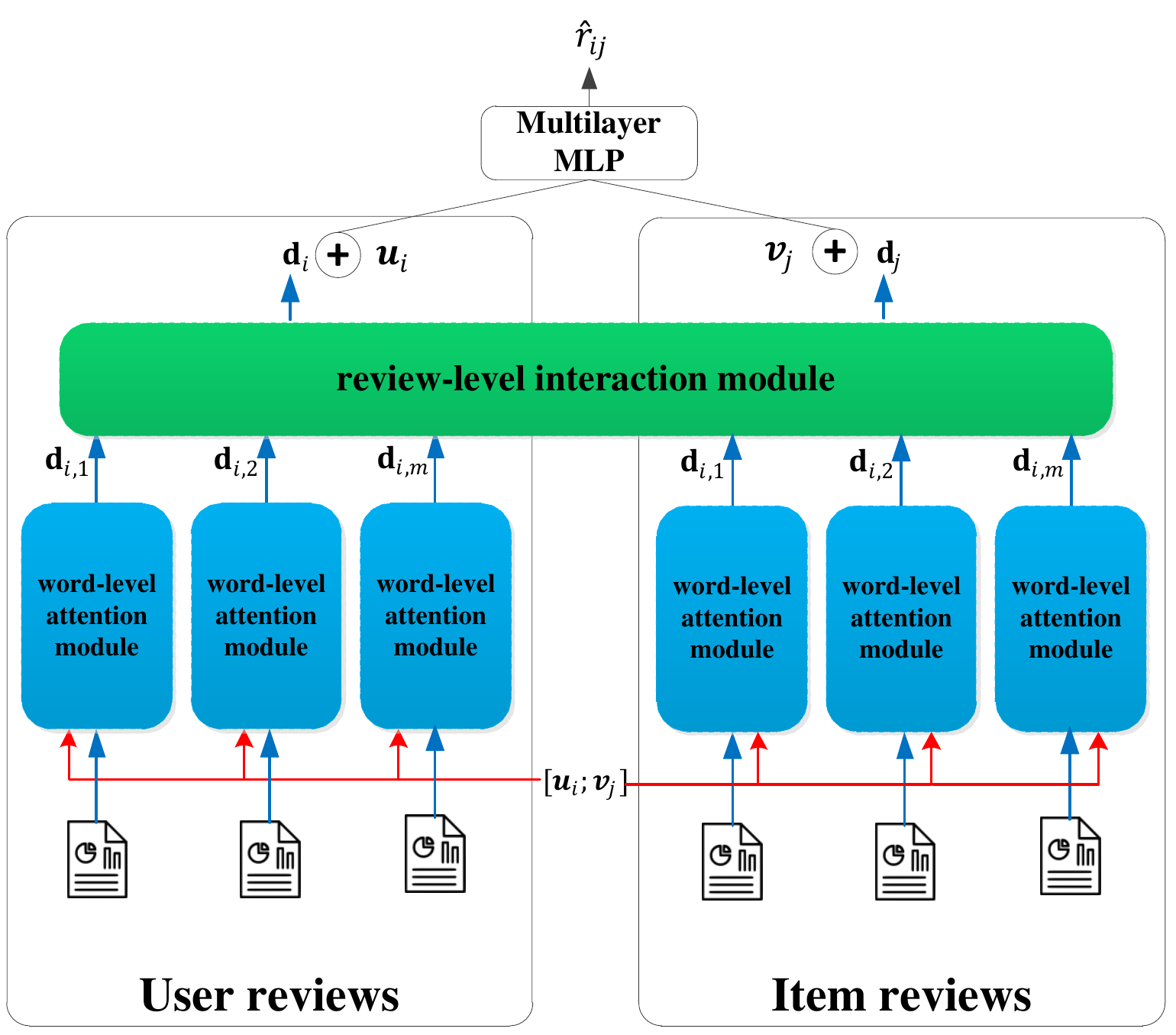}
  }
  \subcaptionbox{(b)Word-level attention module\label{subfig:word}}[.59\linewidth]
  {%
    \includegraphics[width=.9\linewidth]{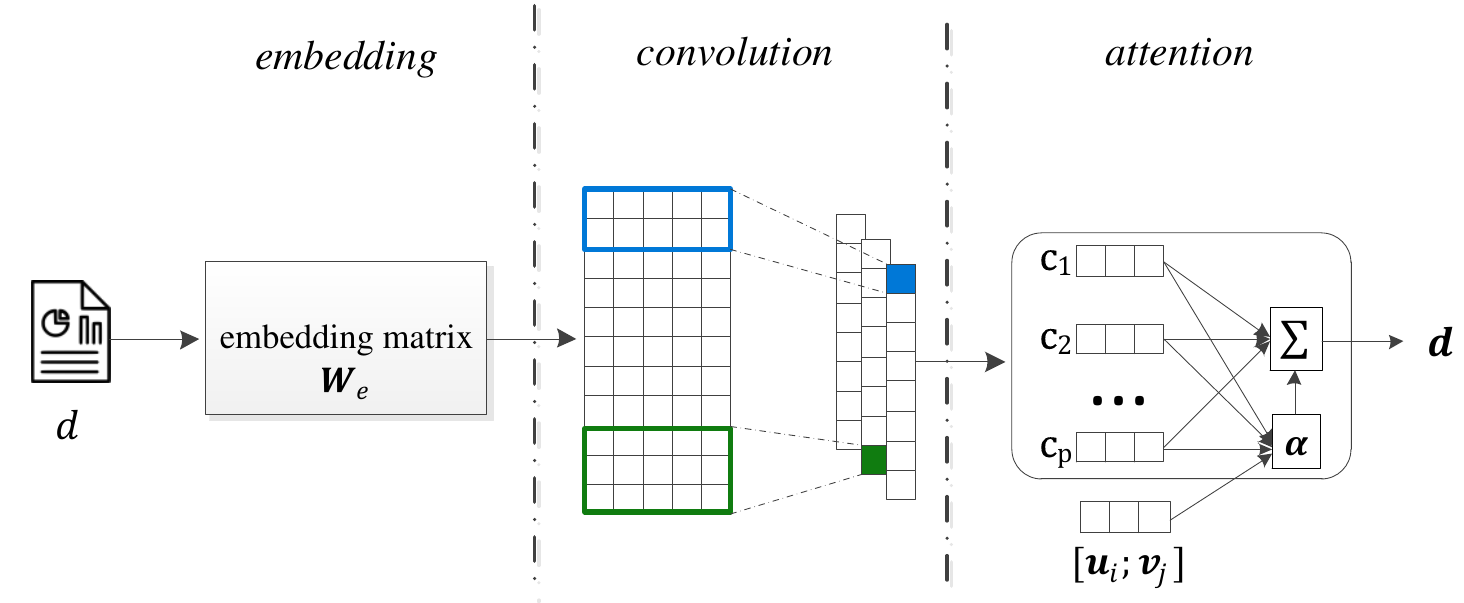}}
  \subcaptionbox{(c)Review-level interaction module\label{subfig:review}}[.39\linewidth]
  {%
    \includegraphics[width=.8\linewidth]{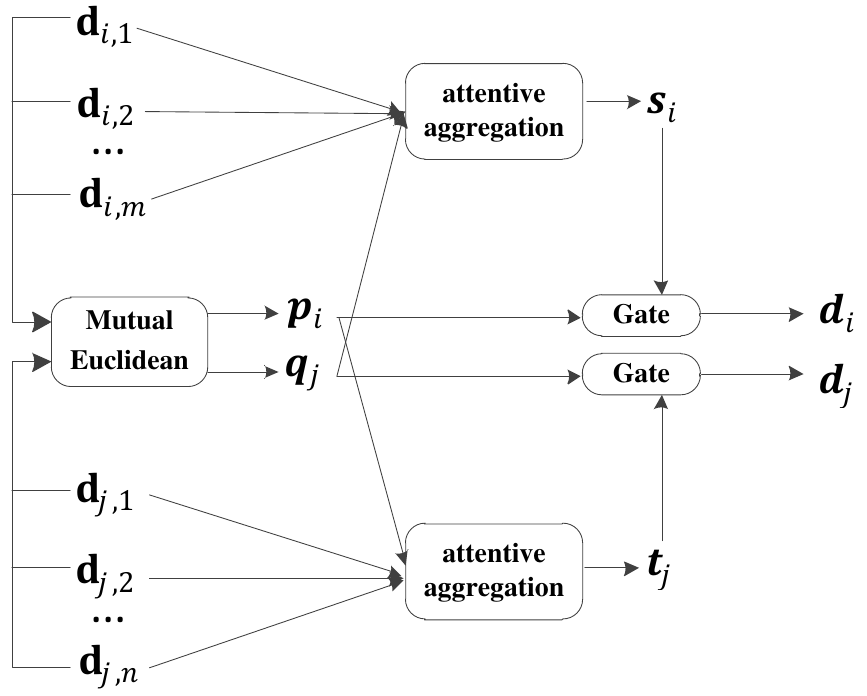}}

  \caption{(a) shows the overall architecture of the proposed model, which consists of a word-level attention module and a review-level interaction module. The two sub-modules are presented in (b) and (c), respectively.\label{fig:modules} }

\end{figure*}

Fig.\ref{fig:modules}(a) illustrates the architecture of the proposed model, which can be decomposed into two parts, namely word-level attention module and review-level interaction module. The word-level attention module is indicated in the blue rectangle in Fig.\ref{fig:modules}(a). It is applied to each review document for extracting high-level word vectors with CNNs. Then, an attention mechanism is proposed to summarize the word vectors into a review representation (\eg $\mathbf{d}_{i,1}$). The review-level interaction module is illustrated with a green rectangle in Fig.\ref{fig:modules}(a).  It models mutual correlations between user and item reviews, and produces an aggregated representation (i.e. $\mathbf{d}_i,\mathbf{d}_j$) for the user and item reviews respectively. Finally, the aggregated representations are combined with the corresponding latent factors ($i.e. \mathbf{u}_i,\mathbf{v}_j$) for rating prediction. The details of the two modules are depicted in Fig.\ref{fig:modules}(b) and Fig.\ref{fig:modules}(c) respectively.

In the word-level attention module, an attention mechanism is proposed to aggregate the words for each review. The attention scores are parameterized by each user-item pair, so that it is able to identify the pair-wise important words. In the review-level interaction module, an interactive network is proposed to aggregate the user and item reviews. To obtain the attention scores, we first mutually propagate textual features between user and item reviews. After that, we calculate the attention scores based on the relevances between user (item) reviews and the propagated item (user) textual features. Therefore, we can identify the most interrelated reviews for the prediction task. The combination of the two modules results in a hierarchical model, which is able to capture the most correlated textual features at different levels of granularities for rating predictions.

\subsection{Word-level Attention Module}
\label{sec:word-level}
Previous works \cite{liu2019daml, wu2019context, zheng2017joint} concatenate that all the reviews of a user/item into a single review, and apply convolution and maxpooling operations for extracting review representations. On the contrary, we apply CNNs with an attention mechanism to each review, and then aggregate the resulted review representations with an interaction module. The rationale is to preserve textual information in a hierarchical manner for the prediction task.

As shown in Fig.\ref{fig:modules}(b), we first embed the words of a review into a sequence of low-dimension dense embeddings,  and then apply 1-dimension CNNs to extract high word vectors. Finally, we utilize an attention mechanism to summarize those word vectors into a review representation.

\paragraph{Embedding}The embedding operation is to map a word to its corresponding word embedding, and it is achieved by indexing an embedding matrix with its word indices in the vocabulary. Without loss of generality, we denote a review as $d=\{w_i\}_{i=1}^p$, where $p$ is the length of the review. The word sequence in $d$ is then embedded into a sequence of word embeddings, $\{\mathbf{w}_i\}_{i=1}^p$, using a word embedding matrix $\mathbf{W}_e\in\mathbb{R}^{v\times dim}$, where $v$ denotes the vocabulary size, and $dim$ denotes the dimension of the word embeddings.

\paragraph{Convolution}CNNs are widely employed in natural language processing area, as they are proven to be powerful in specific tasks such as text classification \cite{kim2014convolutional}. In CNNs, each filter map performs convolution operation from n-gram word embeddings, and it is slid over the review to capture n-gram local features at different positions of the sequence. When filter maps of different sizes are applied,  they are assumed to be capable of capturing local semantics of different granularities. Specifically, we apply a convolution filter map $\mathbf{W}_f\in\mathbb{R}^{s\times dim}$ of size $s$ over a window of $s$ words centered at $w_i$, and produce a single convolutional feature:
\begin{equation}
c_{i,f} = f([\mathbf{w}_{i-(s-1)/2};\cdots;\mathbf{w}_{i+(s-1)/2}]*\mathbf{W}_f+b_f)
\end{equation}
where $[;]$ is the concatenation operation, $f$ is a non-linear function (e.g. relu function \cite{glorot2011deep}), $b_f$ is a bias scalar and $*$ denotes the convolution operation. In this work, $s$ is an odd number.  In practice, we apply filter maps of various sizes to produce multiple features for each word $w_i$. Those features are concatenated to form a convolutional vector for $w_i$: $\mathbf{c}_i = [c_{i,f_1};c_{i,f_2};\cdots;c_{i,f_k}]$, where $k$ is the number of filter maps, and $c_{i,f_k}$ is the convolutional feature produced by the $k$-th filter map.

\paragraph{Attention}Given the word vectors of a review, we propose a pair-specific attention mechanism to aggregate those vectors into a review representation. The basic idea is that not all words have equal contribution for comprising the review representation. In addition, users may have different preferences over the words when composing the review given different items. For example, ``inexpensive" is informative for users who focus on the price, while ``lightweight" needs to be paid high attention for users who care about the portability of the products.
Therefore, the attention scores need to be specialized by the user-item pairs. Specifically, the review representation can be aggregated as the weighted sum over the word vectors:
\begin{equation}
\mathbf{d} = \sum_{i=1}^p \alpha_i\mathbf{c}_i
\end{equation}
where $\alpha_i$ is the attention score assigned to $\mathbf{c}_i$, and it is proportional to the relevance between the user-item pair and $\mathbf{c}_i$:
\begin{equation}
\begin{split}
&\mathbf{m} = tanh(\mathbf{W}^T[\mathbf{u}_i;\mathbf{v}_j]+\mathbf{b})\\
&\alpha_i = \frac{exp(\mathbf{m}^T\mathbf{c}_i)}{\sum_{i=1}^pexp(\mathbf{m}^T\mathbf{c}_i)}
\end{split}
\end{equation}
where matrix $\mathbf{W}$ and vector $\mathbf{b}$ are model parameters, and $\mathbf{u}_i$ and $\mathbf{v}_j$ are latent factors of $u_i$ and $v_j$ respectively. As illustrated, we first transform the concatenated latent factors of $u_i$ and $v_j$ into a latent vector $\mathbf{m}$, and measure its relevance with the word vectors in the common latent space. To obtain the attention scores, the attention scores are then normalized with a softmax function.

\subsection{Review-level Interaction Module}
%

We denote the set of review representations of user $u_i$ and item $v_j$ as $\{\mathbf{d}_{i,k}\}_{k=1}^m$ and $\{\mathbf{d}_{j,t}\}_{t=1}^n$, where $m$ and $n$ denote the number of reviews of $u_i$ and $v_j$ respectively. Each of the review representation is generated by the word-level attention in parallel, as specified in Section.\ref{sec:word-level}. The essence of this module is a mutual learning mechanism, which enables textual information to propagate between each user-item pair. The propagated information provides guidelines for the user and item to select the most relevant reviews. The rationale is that not all reviews are equally important, and user/item reviews that have high relevance to item/user reviews are more informative \cite{TIST_cite_leizhu2020}. Those reviews need to be paid high attention, as they represent semantic correlations between user preferences and item characteristics.

The dataflow of this module  is illustrated in Fig.\ref{fig:modules}(c), we first calculate the initial representations (i.e. $\mathbf{p}_i, \mathbf{q}_j$) over the user and item reviews, respectively. Then, the initial representations are used to propagate information between users and items, and find the mutually relevant reviews for comprising the intermediate representations (i.e. $\mathbf{s}_{i}, \mathbf{t}_{j} $). Finally, we employ a gate network to balance the initial representations and intermediate representations, and produce the final  representations (i.e. $\mathbf{d}_i, \mathbf{d}_j$) of user and item reviews respectively.



\paragraph{Initial Representations} The initial representations are calculated with an attention mechanism. The attention scores are inversely proportional to the minimum Euclidean distance between each review and all the counterpart reviews. Therefore, reviews with smaller distances are weighted higher for aggregating the initial representations. To determine the textual features initially propagated between each user-item pair, we first calculate the mutual Euclidean distance between the user and item reviews.
\begin{equation}
\label{eq:euc}
\begin{split}
&e_{k,t} = |\mathbf{d}_{i,k}-\mathbf{d}_{j,t}|\\
&k=1,\cdots,m;\quad t=1,\cdots,n
\end{split}
\end{equation}
where $e_{k,t}$ is the Euclidean distance between $k$-th user review and $t$-th item review. For each user/item review, we select the minimum distance to the counterpart reviews. The selection is achieved through a min-pooling operation:
\begin{equation}
\label{eq:euc_min}
\begin{split}
a_k = min(e_{k,1},e_{k,2},\cdots,e_{k,n})\\
b_t = min(e_{1,t},e_{2,t},\cdots,e_{m,t})
\end{split}
\end{equation}
where $a_k$ indicates the closest distance between $k$-th user review and all the item reviews, and $b_t$ indicates the closest distance between $t$-th item review and all the user reviews. Therefore, a review with small $a_k$ is assigned a large attention score for representing the initial representation of user reviews:
\begin{equation}
\label{eq:init_semantic}
\begin{split}
&\delta_{i,k} = \frac{exp(-a_k)}{\sum_{k=1}^m exp(-a_k)}\\
&\mathbf{p}_i = \sum_{k=1}^m\delta_{i,k}\mathbf{d}_{i,k}
\end{split}
\end{equation}

Similarly, the initial representation of item (\ie $v_j$) reviews can be obtained as follows:
\begin{equation}
\label{eq:init_semantic_1}
\begin{split}
&\delta_{j,t} = \frac{exp(-b_t)}{\sum_{t=1}^n exp(-b_t)}\\
&\mathbf{q}_j = \sum_{t=1}^n\delta_{j,t}\mathbf{d}_{j,t}
\end{split}
\end{equation}
where $\mathbf{p}_i$ and $\mathbf{q}_j$ are the initial representations of user (\ie $u_i$) and item  (\ie $v_j$) reviews respectively. $\delta_{i,k}$ and $\delta_{j,t}$ are normalized scalars that are inversely proportional to $a_k$ and $b_t$ respectively. Therefore, a user review $\mathbf{d}_{i,k}$ having smaller $a_k$ receives larger weight $\delta_{i,k}$ for comprising $\mathbf{p}_i$, and the same for each item review.

\paragraph{Intermediate Representations} The initial representations, $\mathbf{p}_i, \mathbf{q}_j$, are then used to propagate textual information between user and item reviews. The propagation is to uncover mutually correlated reviews for rating prediction. The semantic correlations are captured through an attention mechanism, which assigns large attention scores to reviews that are highly relevant to the propagated information.
\begin{equation}
\label{eqn:att1}
\begin{split}
&\gamma_{i,k} = \mathbf{v}_1^T tanh(\mathbf{W}_u^T\mathbf{q}_j+\mathbf{W}_{vu}^T\mathbf{d}_{i,k}+\mathbf{b}_u)\\
&\beta_{i,k} = \frac{exp(\gamma_{i,k})}{\sum_{k=1}^mexp(\gamma_{i,k})}\\
&\mathbf{s}_{i} = \sum_{k=1}^m\beta_{i,k}\mathbf{d}_{i,k}
\end{split}
\end{equation}
where matrices $\mathbf{W}_u,\mathbf{W}_{vu}$ and vectors $\mathbf{v}_1, \mathbf{b}_u$ are model parameters. That is, we first measure the relevance between $\mathbf{q}_j$ and each of the user review $\mathbf{d}_{i,k}$, and obtain a relevance score $\gamma_{i,k}$ for each user review $\mathbf{d}_{i,k}$. After that, we use the normalized relevance scores $\{\beta_{i,k}\}_{k=1}^m$ to aggregate $\{\mathbf{d}_{i,k}\}_{k=1}^m$ into an intermediate representation $\mathbf{s}_{i}$. As shown in the equation, the attention over the user reviews is guided by the initial representation of item reviews. The underlying rationale is to highlight the most interrelated user reviews that can possibly meet the item characteristics.

Similarly, we use $\mathbf{p}_i$ to propagate information from user reviews to item reviews, and select the most correlated reviews to comprise the intermediate representation of item reivews:
\begin{equation}
\label{eqn:att2}
\begin{split}
&\gamma_{j,t} = \mathbf{v}_2^T tanh(\mathbf{W}_v^T\mathbf{p}_{i}+\mathbf{W}_{uv}^T\mathbf{d}_{j,t}+\mathbf{b}_v)\\
&\beta_{j,t} = \frac{exp(\gamma_{j,t})}{\sum_{t=1}^nexp(\gamma_{j,t})}\\
&\mathbf{t}_j = \sum_{t=1}^n\beta_{j,t}\mathbf{d}_{j,t}
\end{split}
\end{equation}
where matrices $\mathbf{W}_v, \mathbf{W}_{uv}$ and vectors $\mathbf{v}_2,\mathbf{b}_v$ are model parameters, $\gamma_{j,t}$ is the relevance score between $\mathbf{p}_i$ and $\mathbf{d}_{j,t}$ and $\beta_{j,t}$ is the softmax normalization of $\gamma_{j,t}$.

\paragraph{Aggregated Representations} Based on the initial representations (e.g. $\mathbf{p}_i,\mathbf{q}_j$) and the intermediate representations (e.g. $\mathbf{s}_{i}, \mathbf{t}_{j}$), we produce the aggregated representations for the user and item reviews. As we do not know how each user/item balances these two representations, we employ a gate network to automatically model the combination of these two representations as follows.
\begin{equation}
\label{eqn:gat1}
\begin{split}
&\mathbf{g}_u = \sigma(\mathbf{W}_{g1}^T\mathbf{p}_i+\mathbf{W}_{g2}^T\mathbf{s}_{i}+\mathbf{b}_g)\\
&\mathbf{d}_i = \mathbf{g}_u\odot\mathbf{p}_i+(1-\mathbf{g}_u)\odot\mathbf{s}_{i}
\end{split}
\end{equation}
where matrices $\mathbf{W}_{g1}, \mathbf{W}_{g2}$ and vector $\mathbf{b}_g$ are model parameters, $\sigma(x)=\frac{1}{1+exp(-x)}$ is the sigmoid function, and $\odot$ denotes the element-wise product. Similarly, the gate layer to model the balance between $\mathbf{q}_j$ and $\mathbf{t}_{j}$ is formulated as follows,

\begin{equation}
\label{eqn:gat2}
\begin{split}
&\mathbf{g}_v = \sigma(\mathbf{W}_{g1}^T\mathbf{q}_j+\mathbf{W}_{g2}^T\mathbf{t}_{j}+\mathbf{b}_g)\\
&\mathbf{d}_j = \mathbf{g}_v\odot\mathbf{q}_j+(1-\mathbf{g}_v)\odot\mathbf{t}_{j}
\end{split}
\end{equation}
where the model parameters $\mathbf{W}_{g1}, \mathbf{W}_{g2}, \mathbf{b}_g$ are shared across user and item gate networks.

\subsection{Prediction Module}
\label{sec:prediction}
Notice that the aggregated representations of user and item reviews capture the explicit textual correlations between the users and items, while the latent factors implicitly model the rating behaviours encoded in the rating matrix $\mathbf{R}$. Therefore, we propose to seamlessly integrate the aggregated representations and latent factors with nonlinear transformations. Inspired by \cite{chen2018neural}, given a user-item pair ($u_i,v_j$), we first linearly combine the latent factors and aggregated representations for the user and item:
\begin{equation}
\mathbf{x}_i = \mathbf{u}_i+\mathbf{d}_i;\quad \mathbf{y}_j = \mathbf{v}_j+\mathbf{d}_j
\end{equation}
where $\mathbf{u}_i$ and $\mathbf{v}_j$ are latent factors of $u_i$ and $v_j$. The combined user and item vectors are then input to multilayer neural networks to estimate the rating score:
\begin{equation}
\begin{split}
&\mathbf{h}_0 = [\mathbf{x}_i;\mathbf{y}_j;\mathbf{x}_i\odot\mathbf{y}_j]\\
&\hat{r}_{ij} = \phi_L(...\phi_2(\phi_1(\mathbf{h}_0))...)\\
&\phi_l = \sigma_l(\mathbf{W}_l^T\mathbf{h}_{l-1}+\mathbf{b}_l)\\
\end{split}
\end{equation}
where $\phi_l$ indicates the $l$-th layer neural network, and $\sigma_l, \mathbf{W}_{l}, \mathbf{b}$ are the respective activation function, weight matrix and weight vector. As we focus on explicit rating prediction, we do not apply nonlinear transformations to the last-layer neural network.

\subsection{Model Learning}
We formulate the learning problem as a regression problem. Therefore, the objective function of the proposed model can be defined as the square loss  \cite{xiao2017attentional} \cite{li2017neural} between the predicted and real ratings:
\begin{equation}
\mathcal{L} = \frac{1}{|\mathcal{D}|}\sum_{(u_i,v_j,r_{ij})\in \mathcal{D}}(r_{ij}-\hat{r}_{ij})^2+\lambda||\Theta||^2
\end{equation}
where $\mathcal{D}$ denotes the training set, $r_{ij}$ is the ground truth rating and $\hat{r}_{ij}$ is the predicted rating, $\Theta$ denotes all the the model parameters and $\lambda$ is the regularization tradeoff. The term $\lambda||\Theta||^2$ is introduced to avoid overfitting. We optimize the object function with Adam optimizer \cite{kingma2014adam}, which is a variant of Stochastic Gradient Descent with a dynamically tuned learning rate. The model parameters are updated at every step along the gradient direction with the following protocol:
\begin{equation}
\label{eq:learn}
\Theta^t\leftarrow\Theta^{t-1}-\eta\frac{\partial\mathcal{L}}{\partial\Theta}
\end{equation}
where $\eta$ is the learning rate, and $\frac{\partial\mathcal{L}}{\partial\Theta}$ is the set of partial derivatives of the object function with respect to the model parameters. Those partial derivatives can be automatically computed with typical deep learning libraries. The algorithm of the forward pass and backpropagation of the proposed model are summarized in Algorithm \ref{alg:model}. The presented algorithm processes a  user-item record for each iteration. However, in practice, it is can be easily extended to process a batch of data per iteration. Line 4-6 correspond to word-level attention module for comprising review representations. Line 7-9 specify review-level interaction module for aggregating user and item reviews. In line 10-11, we calculate the predicted rating and back propagate the loss to optimize the model parameters.

 \begin{algorithm}\footnotesize
 \caption{forward pass and back-propagation learning of HTI.}
 \label{alg:model}
 \begin{algorithmic}[1]
   \Require
   the training set $\mathcal{D}$; the learning rate $\eta$; the regularization parameter $\lambda$;
   the reviews for each user $u_i$ and item $v_j$, $D_{u_i}$ and $D_{v_j}$; the number of interactive gate layer, $l_g$
   \Ensure
   latent factors of each user $u_i$ and item $v_j$, $\mathbf{u}_i$ and $\mathbf{v}_j$; aggregated representations of user and item reviews, $\mathbf{d}_i$ and $\mathbf{d}_j$; the model parameters,$\Theta$.
   \State Initialize all model parameters $\Theta$;
   \While{not convergence}
   	\State Randomly sample a tuple $(u_i,v_j,r_{ij})\in\mathcal{D}$;
   	\State Obtain the user and item embeddings;
   	\State Obtain word embeddings for each review in $D_{u_i}$ and $D_{v_j}$;
   	\State Calculate review representations $\{\mathbf{d}_{i,k}\}_{k=1}^m$ and $\{\mathbf{d}_{j,t}\}_{t=1}^n$  as specified in Section.\ref{sec:word-level};
   	\State Calculate initial representations of the user and item reviews, $\mathbf{p}_i$ and $\mathbf{q}_j$ by Eqn.\ref{eq:euc}, Eqn.\ref{eq:euc_min}, Eqn.\ref{eq:init_semantic} and Eqn.\ref{eq:init_semantic_1};
   	\State Calculate intermediate representations of user and item reviews, $\mathbf{s}_i$ and $\mathbf{t}_j$, as specified in Eqn.\ref{eqn:att1} and Eqn.\ref{eqn:att2}.
   	\State Calculate aggregated representations of user and item reviews, $\mathbf{d}_i$ and $\mathbf{d}_j$, as specified in Eqn.\ref{eqn:gat1} and Eqn.\ref{eqn:gat2}.
   	\State Calculate the predicted score $\hat{r}_{ij}$ as descried in Section.\ref{sec:prediction};
   	\State Calculate $\frac{\partial\mathcal{L}}{\partial\Theta}$ and update $\Theta$ by Eqn.\ref{eq:learn};
   \EndWhile
 \end{algorithmic}
 \end{algorithm}

\subsection{Time Complexity Analysis}
\label{sec:time}
For a given user-item pair, the time complexity of a forward pass of the proposed model can be decomposed into the following parts. The time complexity of the convolution layer is $O(dim*k*l_d*(m+n))$, where $dim$ is the embedding size, $k$ is the number of convolutional features (i.e. number of feature maps), $l_d$ denotes the maximum review length, and $m$ and $n$ denote the number of user and item reviews, respectively. In the word-level attention layer, the time complexity to calculate the attention scores is $O(k^2(m+n))$. The time complexity of review-level interaction network is $O(mnk+k^2(m+n))$, where the first term is the time complexity for computing the initial representations, while the second term is the time complexity for calculating the aggregated representations. In the prediction layer, the operations of the multi-layer neural networks can be completed in $O(\sum_{l=1}^Lk_{l}k_{l+1})$, where $k_l$ is the dimension of the $l$-th layer. In reality, $m,n,l_i,L$ are much smaller than $dim$ and $k$, hence the time complexity of the proposed model mainly depends on the embedding size and quantity of latent features.
%
%

%

\section{Experiments}
\subsection{Datasets}
\begin{table*}
\centering
\caption{\normalsize{Statistics of the datasets, where \#doc/user and \#doc/item indicate the average number of reviews per user and item respectively, and \#word/doc is the average number of words per review.}}
\begin{tabular}{llllllll}
\toprule
\textbf{Dataset} & \textbf{\#user} &\textbf{\#item} & \textbf{\#rating} & \textbf{\#doc/user} & \textbf{\#doc/item} & \textbf{\#word/doc} & \textbf{Density}\\
\hline
Musical Instruments & 1,429 & 900  & 10,261 & 10 & 17 & 59 & 0.798\%\\
Office Products & 4905 & 2420 & 53,228 & 7 & 11 & 56 &  0.449\%\\
Grocery and Gourmet Food & 14,681 & 8,713 & 151,254 & 10 & 22 & 89 & 0.118\%\\
Video Games & 24,303 & 10,672 & 231,577 & 9 & 21 & 124 & 0.089\%\\
Sports and Outdoors & 35,598 & 18,357 & 296,337 &8 &14 &99& 0.045\%\\
\bottomrule
\end{tabular}
\label{tb:stats}
\end{table*}
To validate the effectiveness of the proposed model, we conduct experiments on five publicly available datasets. The five datasets are from different domains of Amazon 5-core \cite{he2016ups}: Music Instruments, Office Products, Grocery and Gourmet Food, Video Games, Sports and Outdoors. Following the common protocol in \cite{wu2019context}, we first process all the characters into the lower cases, and filter out the stop words. After that, we select the 20,000 most frequent words as vocabulary, and remove all the words out of the vocabulary from raw reviews. Finally, following the previous protocol \cite{chen2018neural}, we pad the length and the number of reviews to cover  90\% users and items respectively. Following previous work \cite{catherine2017transnets}, for a  user-item pair, we remove the review that the user writes about the item from the review repositories of both of the user and item.

The statistics of the datasets are shown in Table.\ref{tb:stats}. In comparison, previous works \cite{liu2019daml, wu2019context} concatenate all the reviews and represent each user/item with about a hundred words, while we organize the reviews into a review-word hierarchy. Therefore, we are able to preserve rich textual information at different levels of granularities for accurate rating prediction.

\subsection{Baselines}
To demonstrate the advantage of the proposed model, we compare it with the following state-of-the-art model.
\begin{itemize}
\item PMF \cite{mnih2008probabilistic} is a standard matrix factorization that mainly relies on user-item rating for prediction.
\item NeuMF \cite{he2017neural} integrates generalized matrix factorization and multi-layer perceptron for user-item modeling.
\item CDL \cite{wang2015collaborative} utilizes SDAE \cite{vincent2010stacked} to extract textual features, and hierarchical Beyeisan model to integrate ratings and reviews for rating prediction.
\item ConvMF \cite{kim2016convolutional} leverages CNNs to learn textual features from reviews, and incorporates them into PMF for rating prediction.
\item DeepCoNN \cite{zheng2017joint} extracts user and item features from their respective reviews, and uses factorization machine to derive the prediction.
\item D-attn \cite{seo2017interpretable} proposes dual attention mechanism to achieve the interpretability of latent features of user and item.
\item CARL \cite{wu2019context} applies CNNs to learn textual features from reviews, and a dynamic linear fusion mechanism for obtaining the final rating scores.
\item NARRE \cite{chen2018neural} leverages two parallel CNNs and an attention mechanism to explore the usefulness of the reviews for rating prediction.
\item DAML \cite{liu2019daml} employs local and mutual attention of CNNs to learn features from reviews, and integrates them with latent factor model for predicting ratings.
\end{itemize}

\subsection{Setup}
We tune the hyper-parameters of the proposed model in the validation set. Specifically, we initialize the word embeddings with 100-dimension Glove embeddings \cite{pennington2014glove}. As for the convolution layer, we utilize two-layer CNNs for extracting word vectors. For the first layer CNN, the kernel sizes are set to 3 and 5, and the number of filter maps associated with each kernel are varied amongst [25,150]. As for the second layer CNN, we use a kernel of size 5, and set the number of filter maps equal to the dimension of latent factors. In the prediction module, we employ two hidden layers and one output layer, and dimension of each hidden layer is halved from the previous layer. To prevent overfitting, we deploy dropout \cite{srivastava2014dropout} layers upon all non-linear transformations with a dropout rate of 0.5. The implemented model is trained via stochastic gradient descent over shuffled mini-batches with a batch size of 128. The defined objective function is optimized using Adam \cite{kingma2014adam} optimizer with an initial learning rate of 0.0001. All the experiments and training are done using a NVIDIA GeForce GTX 1070 graphics card with 8G memory. To facilitate community research, the implementation of the proposed model is made publicly available from here \footnotemark[1].
\footnotetext[1]{https://github.com/uqjwen/HIGAN}

In the experiment, we randomly split each dataset into training set, validation set (10\%), and test set (10\%). We repeat the experiments for 10 times to avoid splitting bias, and report the average results over the 10 runs. However, we notice very minimal differences among the performances of different runs. Also, to study the influence of training ratio, we vary the percentage of training set amontst [80\%, 60\%, 40\%].  We use two standard metrics for rating predictions, namely: Mean Absolute Error (MAE) and Root Mean Square Error (RMSE) as shown below.
\begin{equation}
\begin{split}
MAE &= \frac{1}{|\mathcal{T}|}\sum_{(u_i,v_j,r_{ij})\in \mathcal{T}}|r_{ij}-\hat{r}_{ij}| \\
RMSE &= \sqrt{\frac{1}{|\mathcal{T}|}\sum_{(u_i,v_j,r_{ij})\in \mathcal{T}} (r_{ij}-\hat{r}_{ij})^2}
\end{split}
\end{equation}
where $\mathcal{T}$ denotes the test set.

\subsection{Performance Comparison}

%
\begin{table*}\footnotesize
\centering
\caption{\normalsize{ Performance comparison across the dataset for all models (\textbf{80\% training ratio}). The best and the second-best results are highlighted by boldface and underlined respectively.}}
\begin{tabular}{p{1.6cm}|c|cccccccccc}
\toprule
Datasets & Metrics & PMF&NeuMF&CDL&ConvMF&DeepCoNN&D-attn&NARRE&CARL&DAML&HTI \\
\hline

\multirow{2}{2cm}{\textit{Musical Instruments}} & MAE & 1.137&0.72&0.834&0.786&0.759&0.742&0.695&0.677&\underline{0.651}&\textbf{0.611}\\
& RMSE & 1.352&0.904&1.080&1.026&1.003&0.956&0.922&0.878&\underline{0.848}&\textbf{0.813}\\
\hline
\multirow{2}{2cm}{\textit{Office Products}} & MAE & 1.265&0.73&1.062&0.728&0.711&0.716&0.681&0.647&\underline{0.612}&\textbf{0.552}\\
& RMSE & 1.430&0.921&1.223&0.952&0.901&0.923&0.867&0.834&\underline{0.811}&\textbf{0.731}\\
\hline
\multirow{2}{2cm}{\textit{Grocery and Gourmet Food}} & MAE & 1.397&0.943&0.967&0.863&0.802&0.824&0.747&0.753&\underline{0.735}&\textbf{0.672}\\
& RMSE &1.572&1.197&1.190&1.092&1.036&1.070&0.963&0.961&\underline{0.938}&\textbf{0.877}\\
\hline
\multirow{2}{2cm}{\textit{Video Games}} & MAE & 1.395&0.87&0.902&0.899&0.875&0.842&0.799&0.798&\underline{0.788}&\textbf{0.731}\\
& RMSE & 1.606&1.103&1.165&1.145&1.168&1.062&1.039&\underline{1.029}&1.045&\textbf{0.966}\\
\hline
\multirow{2}{2cm}{\textit{Sports and Outdoors}} & MAE & 1.203&0.752&0.852&0.824&0.719&0.784&0.69&0.686&\underline{0.667}&\textbf{0.629}\\
& RMSE &1.376&0.979&1.089&1.013&0.885&0.997&\underline{0.882}&0.888&0.883&\textbf{0.826}\\
\bottomrule
\end{tabular}
\label{tb:cmp1}
\end{table*}

\begin{table*}\footnotesize
\centering
\caption{\normalsize{Performance comparison across the dataset for all models (\textbf{60\% training ratio}). The best and the second-best results are highlighted by boldface and underlined respectively.}}
\begin{tabular}{p{1.6cm}|c|cccccccccc}
\toprule
Datasets & Metrics & PMF&NeuMF&CDL&ConvMF&DeepCoNN&D-attn&NARRE&CARL&DAML&HTI \\
\hline

\multirow{2}{2cm}{\textit{Musical Instruments}} & MAE & 1.160 & 0.751 & 0.860 & 0.805 & 0.778 & 0.767 & 0.709 & 0.695 & \underline{0.670} & \textbf{0.623} \\
& RMSE & 1.377 & 0.928 & 1.103 & 1.035 & 1.028 & 0.973 & 0.928 & 0.898 & \underline{0.864} & \textbf{0.826} \\
\hline

\multirow{2}{2cm}{\textit{Office Products}} & MAE & 1.281 & 0.743 & 1.100 & 0.751 & 0.735 & 0.737 & 0.681 & 0.660 & \underline{0.631} & \textbf{0.573} \\
& RMSE & 1.454 & 0.943 & 1.261 & 0.999 & 0.932 & 0.963 & 0.885 & 0.860 & \underline{0.835} & \textbf{0.747} \\
\hline

\multirow{2}{2cm}{\textit{Grocery and Gourmet Food}} & MAE & 1.415 & 0.966 & 0.984 & 0.885 & 0.819 & 0.841 & 0.760 & 0.785 & \underline{0.755} & \textbf{0.691} \\
& RMSE & 1.590 & 1.216 & 1.232 & 1.117 & 1.049 & 1.094 & 0.996 & 0.978 & \underline{0.954} & \textbf{0.900} \\
\hline

\multirow{2}{2cm}{\textit{Video Games}} & MAE & 1.438 & 0.908 & 0.919 & 0.914 & 0.892 & 0.860 & 0.807 & 0.816 & \underline{0.802} & \textbf{0.742} \\
& RMSE & 1.679 & 1.129 & 1.191 & 1.156 & 1.214 & 1.090 & 1.059 & \underline{1.054} & 1.061 & \textbf{0.986} \\
\hline

\multirow{2}{2cm}{\textit{Sports and Outdoors}} & MAE & 1.227 & 0.762 & 0.878 & 0.832 & 0.724 & 0.822 & 0.695 & 0.696 & \underline{0.695} & \textbf{0.641} \\
& RMSE & 1.400 & 1.005 & 1.124 & 1.036 & 0.908 & 1.039 & \underline{0.890} & 0.898 & 0.902 & \textbf{0.847} \\

\bottomrule
\end{tabular}
\label{tb:cmp2}
\end{table*}

\begin{table*}\footnotesize
\centering
\caption{\normalsize{Performance comparison across the dataset for all models (\textbf{40\% training ratio}). The best and the second-best results are highlighted by boldface and underlined respectively.}}
\begin{tabular}{p{1.6cm}|c|cccccccccc}
\toprule
Datasets & Metrics & PMF&NeuMF&CDL&ConvMF&DeepCoNN&D-attn&NARRE&CARL&DAML&HTI \\
\hline

\multirow{2}{2cm}{\textit{Musical Instruments}} & MAE & 1.181 & 0.749 & 0.882 & 0.817 & 0.793 & 0.778 & 0.721 & 0.704 & \underline{0.680} & \textbf{0.633}\\
& RMSE & 1.434 & 0.952 & 1.157 & 1.075 & 1.047 & 0.999 & 0.962 & 0.919 & \underline{0.886} & \textbf{0.851}\\
\hline
\multirow{2}{2cm}{\textit{Office Products}} & MAE & 1.317 & 0.767 & 1.107 & 0.758 & 0.740 & 0.742 & 0.696 & 0.675 & \underline{0.637} & \textbf{0.579}\\
& RMSE & 1.493 & 0.953 & 1.265 & 0.983 & 0.948 & 0.975 & 0.907 & 0.884 & \underline{0.847} & \textbf{0.767}\\
\hline
\multirow{2}{2cm}{\textit{Grocery and Gourmet Food}} & MAE & 1.467 & 0.991 & 1.003 & 0.907 & 0.843 & 0.852 & \underline{0.778} & 0.789 & 0.790 & \textbf{0.701}\\
& RMSE & 1.693 & 1.271 & 1.243 & 1.170 & 1.085 & 1.100 & 1.016 & 1.023 & \underline{0.992} & \textbf{0.916}\\
\hline
\multirow{2}{2cm}{\textit{Video Games}} & MAE & 1.468 & 0.905 & 0.944 & 0.941 & 0.904 & 0.885 & 0.845 & 0.840 & \underline{0.821} & \textbf{0.780}\\
& RMSE & 1.693 & 1.150 & 1.205 & 1.194 & 1.199 & 1.117 & 1.099 & \underline{1.085} & 1.092 & \textbf{1.005}\\
\hline
\multirow{2}{2cm}{\textit{Sports and Outdoors}} & MAE & 1.255 & 0.782 & 0.884 & 0.851 & 0.766 & 0.820 & 0.718 & 0.721 & \underline{0.705} & \textbf{0.656}\\
& RMSE & 1.427 & 1.011 & 1.166 & 1.052 & 0.930 & 1.044 & 0.930 & 0.934 & \underline{0.925} & \textbf{0.875}\\
\bottomrule
\end{tabular}
\label{tb:cmp3}
\end{table*}

The overall performances of the models across the datasets are presented in Table.\ref{tb:cmp1}, Table.\ref{tb:cmp2} and Table.\ref{tb:cmp3} with different training ratio, from which we have the following conclusions. First, PMF yields the worst performance across the datasets. PMF is mainly based on user-item interactions, hence the performance is limited by the sparsity instinct of the rating matrix in recommendation problems. NeuMF is also the model that relies on the rating data for prediction. However, it outperforms PMF by a large margin on all the datasets, demonstrating the effectiveness of multilayer nonlinear transformations for capturing complex user-item relations. The advantage of nonlinear interactions can also be validated by NeuMF's superiority to CDL across the datasets. Even though CDL exploits textual features with SDAE, it simply incorporates those features into a hierarchical Bayesian model. Therefore, it fails to capture high-order user-item correlations in the recommendation datasets.

As for review-based models, ConvMF achieves better performance than CDL. Since they both utilize shallow network for modeling user-item interactions, the performance improvement can be attributed to the CNNs, which are efficient in capturing local semantics of different levels of granularities. DeepCoNN demonstrates better performance than ConvMF, implying the advantage of factorization machine in capturing high-order nonlinear interactions of features. The performance differences between D-attn and DeepCoNN are data-dependent. Specifically, D-attn presents performance gains over DeepCoNN on \textit{Musical Instruments} and \textit{Video Games}. This portrays the success of the attention mechanism in selecting informative words for rating predictions. However, on \textit{Office Products} and \textit{Grocery and Gourmet Food}, DeepCoNN yields slightly better performance than D-attn, probably because D-attn lacks the expressiveness to capture complex high-order user-item interactions.

NARRE demonstrates improved performance over D-attn, indicating the advantage of exploiting the textual information in a hierarchical manner. D-attn concatenates all the reviews of a user/item into a single review, where the introduced noise and irrelevant information in the reviews can negatively affect the model performance. On the contrary, NARRE extracts textual features for the reviews in parallel, and proposes an attention mechanism to select the most informative reviews for prediction. CARL and DAML share similar model architecture, and they present significant improved performances over NARRE. The comparison results suggest the advantage of modeling the pair-based relevance between users and items, as it can dynamically capture complex interactions for each specific user-item pair.

The proposed HTI model achieves the best MAE/RMSE scores across the five datasets under different training ratios. Specifically, HTI achieves an average improvement of 7.48\% in MAE (7.2\% in RMSE) over the best baseline. Compared with existing models, NARRE is a hierarchical model that exploits both word- and review-level textual information for learning review features. However, at word level,  it takes a max pooling over the word vectors to comprise a review representation. At review level, it utilizes a self-attention mechanism to aggregate the reviews into latent features. One limitation in the max pooling at word level is information loss, namely the most salient feature can override the rest features in the same dimension of a review. In addition, the self-attention mechanism at review level defers the user-item interactions until the prediction layer, making it fail to capture complex user-item correlations. On the contrary, at word level, we propose a pair-specific attention mechanism that can capture informative words for comprising review representations. At review level, we allow textual features to be propagated between each user-item pair. Therefore, we model semantic alignments at review level for modeling complex user-item interactions. The comparison between HTI and previous models such as DAML demonstrates the effectiveness of the proposed hierarchical architecture. Those models simply concatenate all reviews of a user/item into a single review. Therefore, the irrelevant words can introduce noises and undermine the recommendation accuracy. In addition, they model aspect relevance at word/phrase level, which does not necessarily reflect the meaning of the original sentences \cite{ribeiro2016should}. By contrast, we propose to model textual features at both word and review levels, which allow us to capture semantic correlations at different levels of granularities for accurate prediction.

Comparing the recommendation performance of competitive models with different training rations, we can see that the proposed model is able to outperform the baselines under various data sparsity, showing its effectiveness in exploiting hierarchical and interactive textual features for alleviating data sparsity and increasing recommendation performance.


\subsection{Analysis of HTI Model}
\subsubsection{Effect of Two Learning Modules}
%

\begin{table*}\small
\centering
\caption{\normalsize{Comparison results between HTI and its variants. HTI-wavg: replaces the word-level attention layer with an average pooling layer. HTI-wmax: replaces the word-level attention layer with a max pooling layer. HTI-davg: replaces the review-level interactive layer with an average pooling layer. HTI-dmax: replaces the review-level interactive layer with a max pooling layer. * indicates the comparison results are statistically significant at the level of 0.001. }}
\begin{tabular}{lccccc}
\toprule
Variants & \textit{Musical Instruments} & \textit{Office Products} & \textit{Grocery and Gourmet Food} & \textit{Video Games} & \textit{Sports and Outdoors}\\
\hline
HTI-wavg & 0.632 & 0.584 & 0.723 & 0.762 & 0.646\\
HTI-wmax & 0.618 & 0.563 & 0.684 & 0.742 & 0.636\\
HTI-davg & 0.669 & 0.635 & 0.788 & 0.782& 0.690\\
HTI-dmax & 0.654 & 0.619 & 0.760 & 0.793&0.669\\
HTI & 0.611* & 0.552* & 0.672* & 0.731* &0.629*\\
\bottomrule
\end{tabular}
\label{tb:variants}
\end{table*}

To study the effect of the word-level attention module and review-level interaction module, we compare HTI with its variants shown as follows,
\begin{itemize}
\item[\romannumeral1]HTI-wavg obtains each review representation by applying a mean-pooling layer over the word vectors.
\item[\romannumeral2]HTI-wmax obtains each review representation by applying a max-pooling layer over the word vectors.
\item[\romannumeral3]HTI-davg obtains the aggregated representations by applying a mean-pooling layer over the review representations.
\item[\romannumeral4]HTI-dmax obtains the aggregated representations by applying a max-pooling layer over the review representations.
\end{itemize}

The comparison results on the five datasets are presented in Table.\ref{tb:variants}. From the table, we can see that HTI significantly outperforms its variants, demonstrating the effectiveness of modeling text interaction at different levels of granularities for rating prediction. Second, the max pooling layer yields slightly better MAE over the mean pooling layer on the datasets, as max-pooling operation is able to capture the most salient features \cite{xue2018aspect}. Finally, the variants without review-level interactions consistently perform worse than the variants exclude word-level attentions, suggesting the importance of capturing semantic correlations between users and items at the review level. We believe the reasons are two-fold. First, pooling over the review representations does not allow HTI to model text interaction at the review level, making the model unable to capture pair-wise textual alignments and extracting task-specific textual features.  Second, words in a review tend to have similar sentimental polarity, but this is not the case at the review level, as users may write reviews of different polarities for different items. Therefore, pooling reviews of contradictory polarities incurs more recommendation degradation then pooling over words of similar polarities.

The results indicate that the proposed model can effectively preserve textual features at different levels of hierarchies. The significant improvement over HTI-davg and HTI-dmax demonstrates the advantage of exploiting mutual correlations at review level to derive accurate recommendations.
\subsubsection{Number of latent dimension}

\begin{figure}
\centering
\includegraphics[width=7cm]{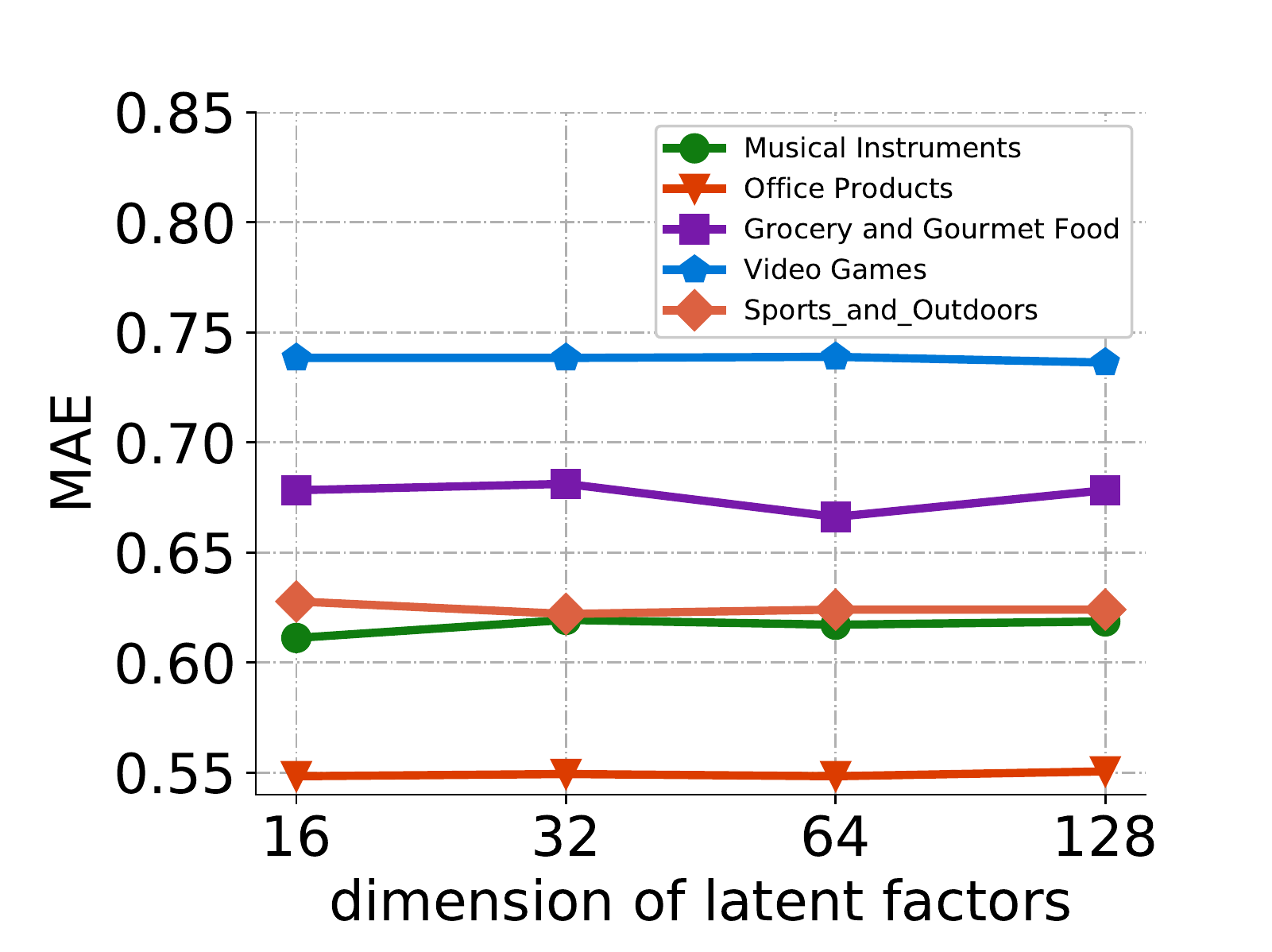}
\caption{Recommendation performance (MAE) with respect to different number of latent dimensions. }
\label{fig:dimension}
\end{figure}
To study the sensitivity of HTI with respect to the number of latent factors, we vary the dimension amongst [16,32,64,128]. As shown in Fig.\ref{fig:dimension}, the hyperparameter has little impact on the model performance, as HTI is able to provide a relatively stable MAE with respect to different dimensions. However, as we analyze in Section.\ref{sec:time}, the time complexity of HTI is related to the latent factor dimensions, a small latent dimension is selected to reduce computational overhead in this work.

\subsubsection{Hyerparameters of CNNs}
\begin{figure*}
  \centering
  \subcaptionbox{\textit{Musical Instruments}}[.32\textwidth][c]{%
    \includegraphics[width=.32\textwidth]{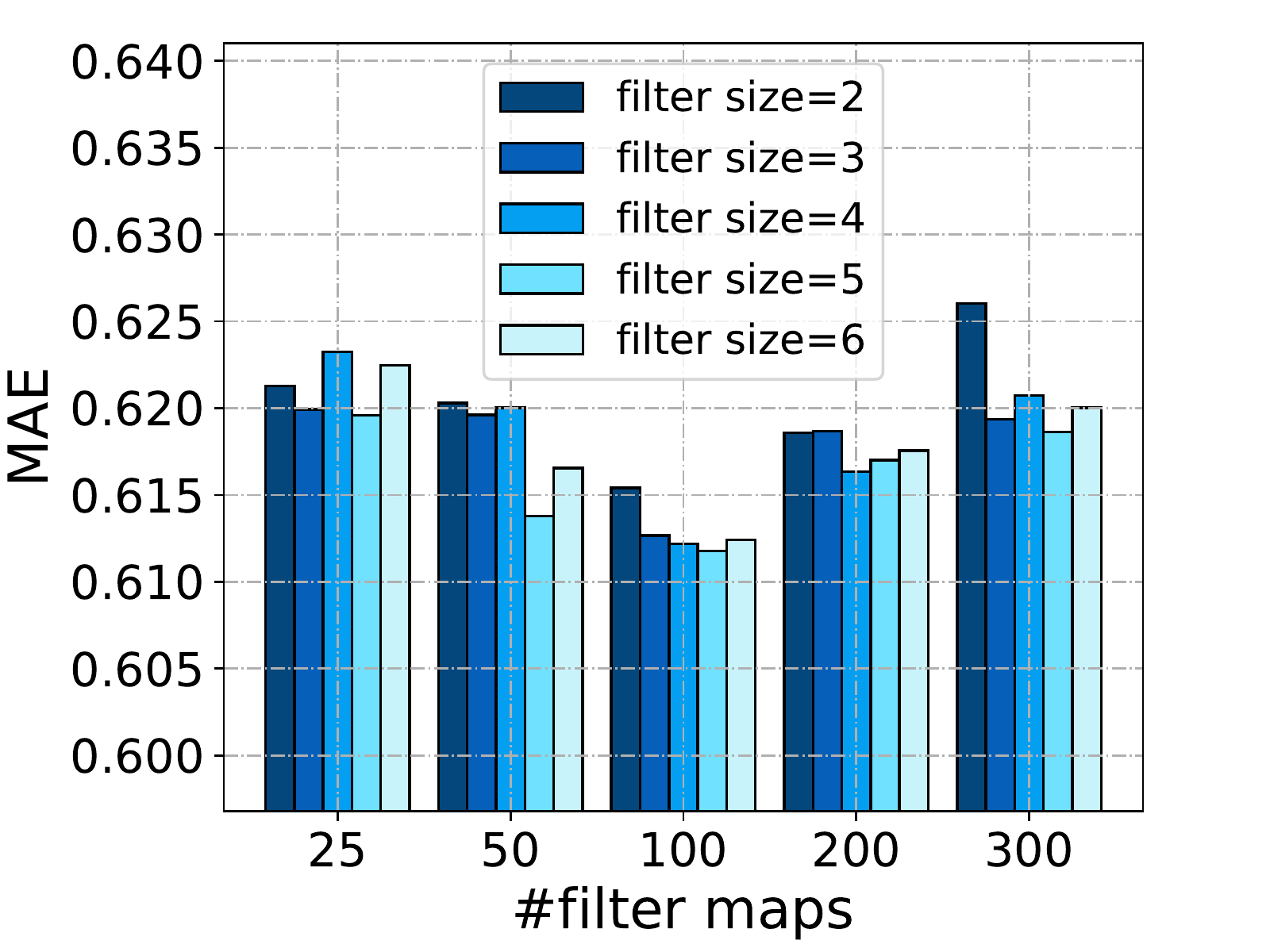}}
  \subcaptionbox{\textit{Office Products}}[.32\textwidth][c]{%
    \includegraphics[width=.32\textwidth]{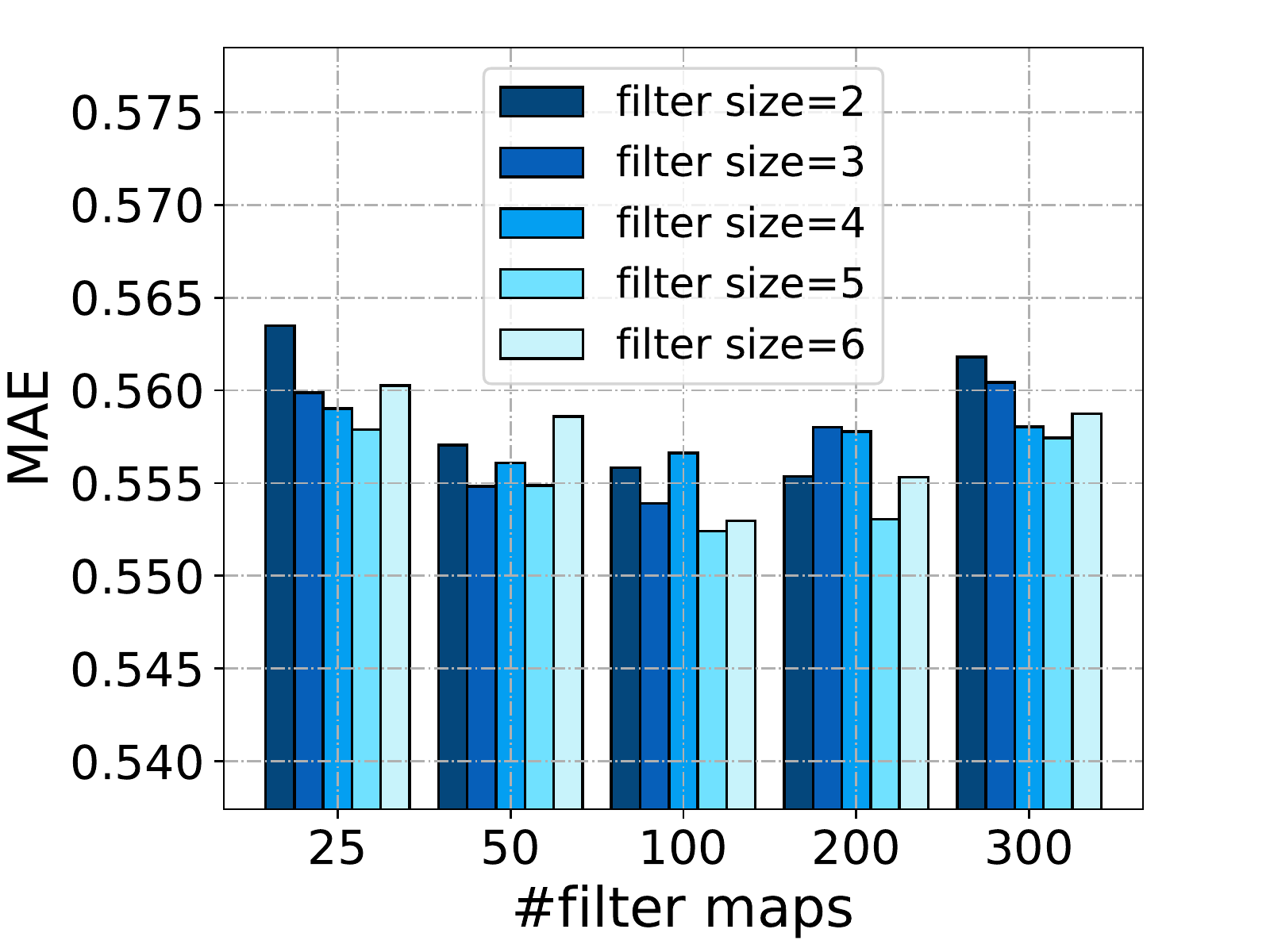}}
  \subcaptionbox{\textit{Grocery and Gourmet Food}}[.32\textwidth][c]{%
    \includegraphics[width=.32\textwidth]{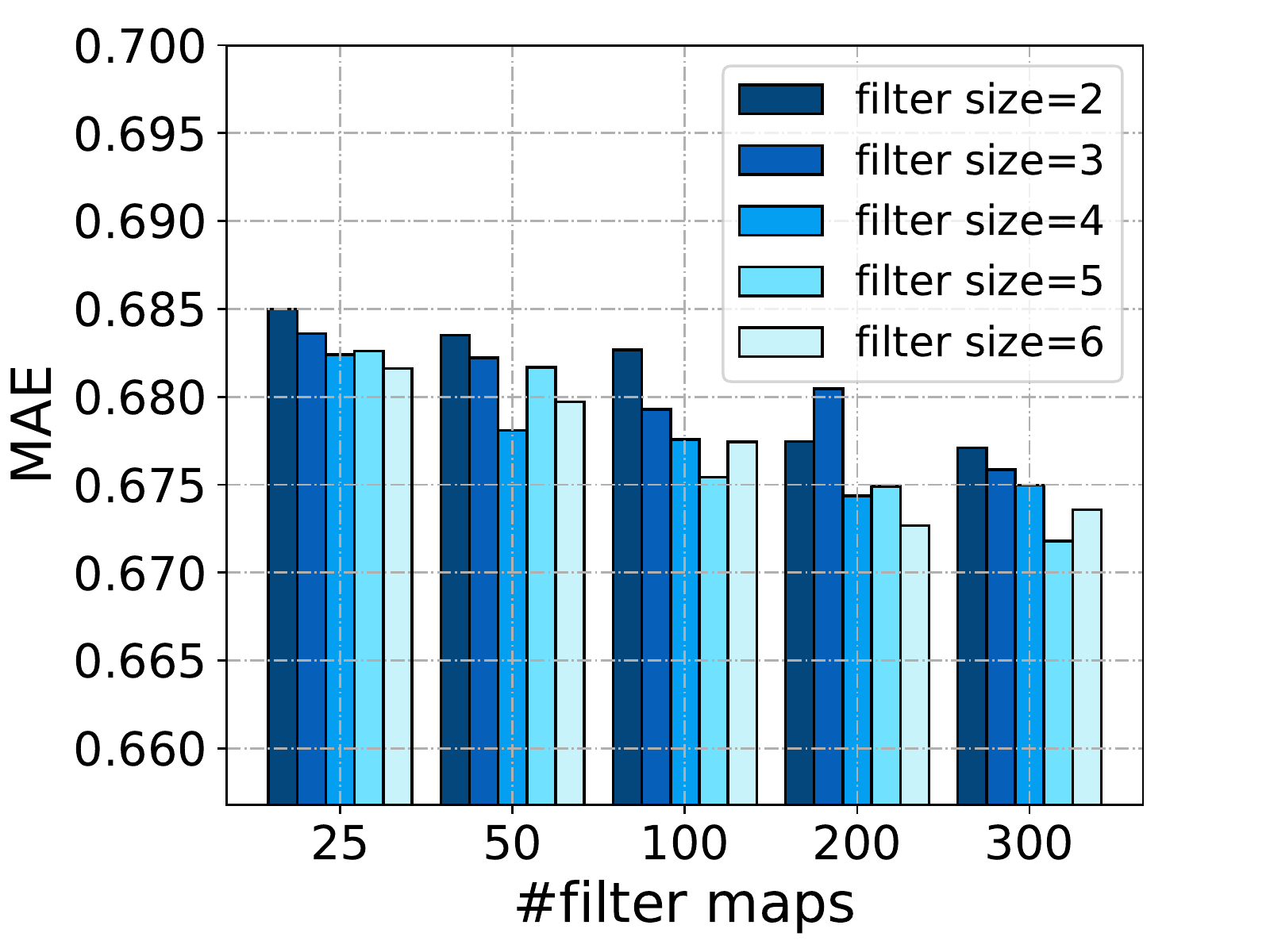}}
  \subcaptionbox{\textit{Video Games}}[.32\textwidth][c]{%
    \includegraphics[width=.32\textwidth]{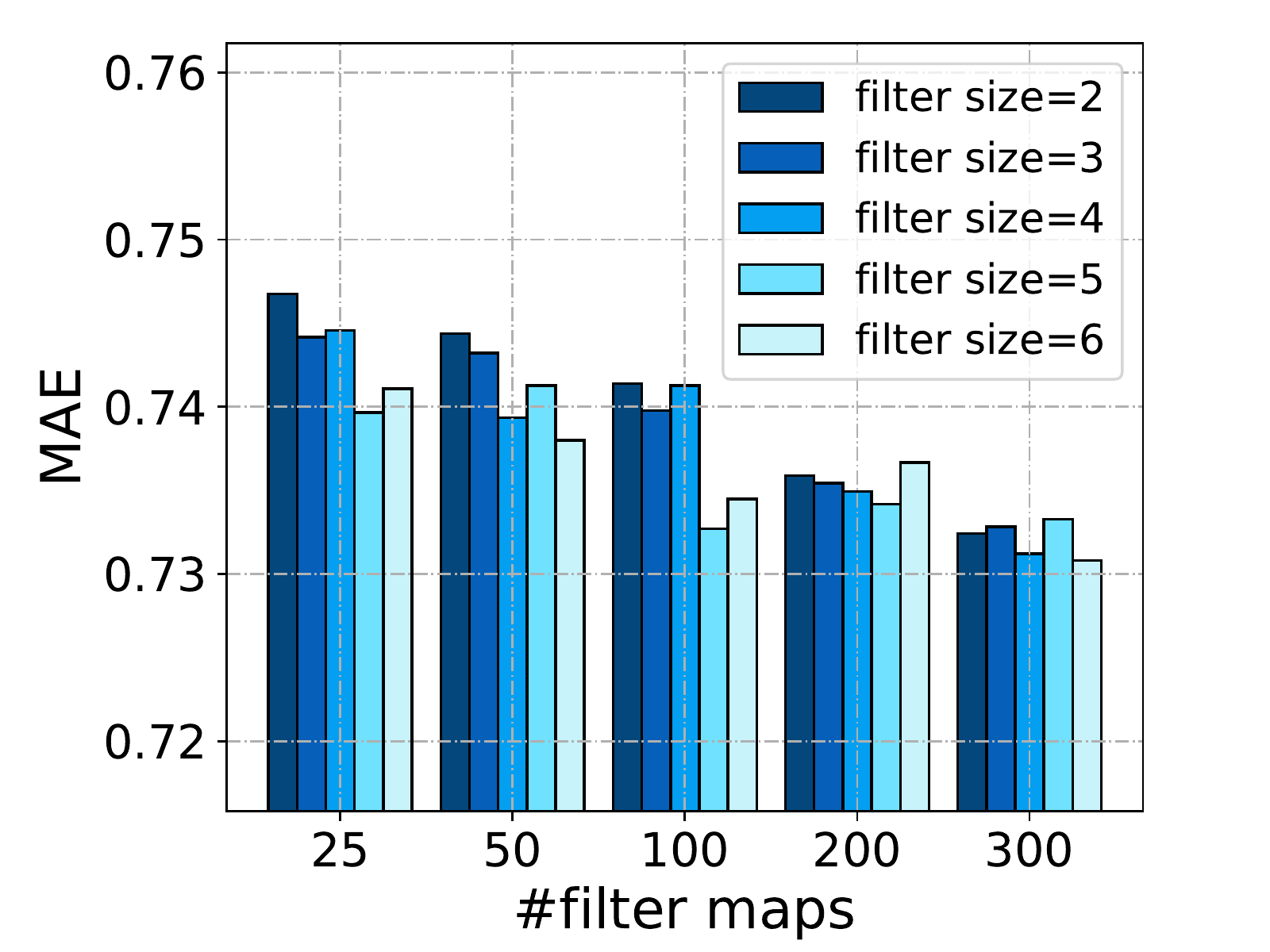}}
  \subcaptionbox{\textit{Sports and Outdoors}}[.32\textwidth][c]{%
    \includegraphics[width=.32\textwidth]{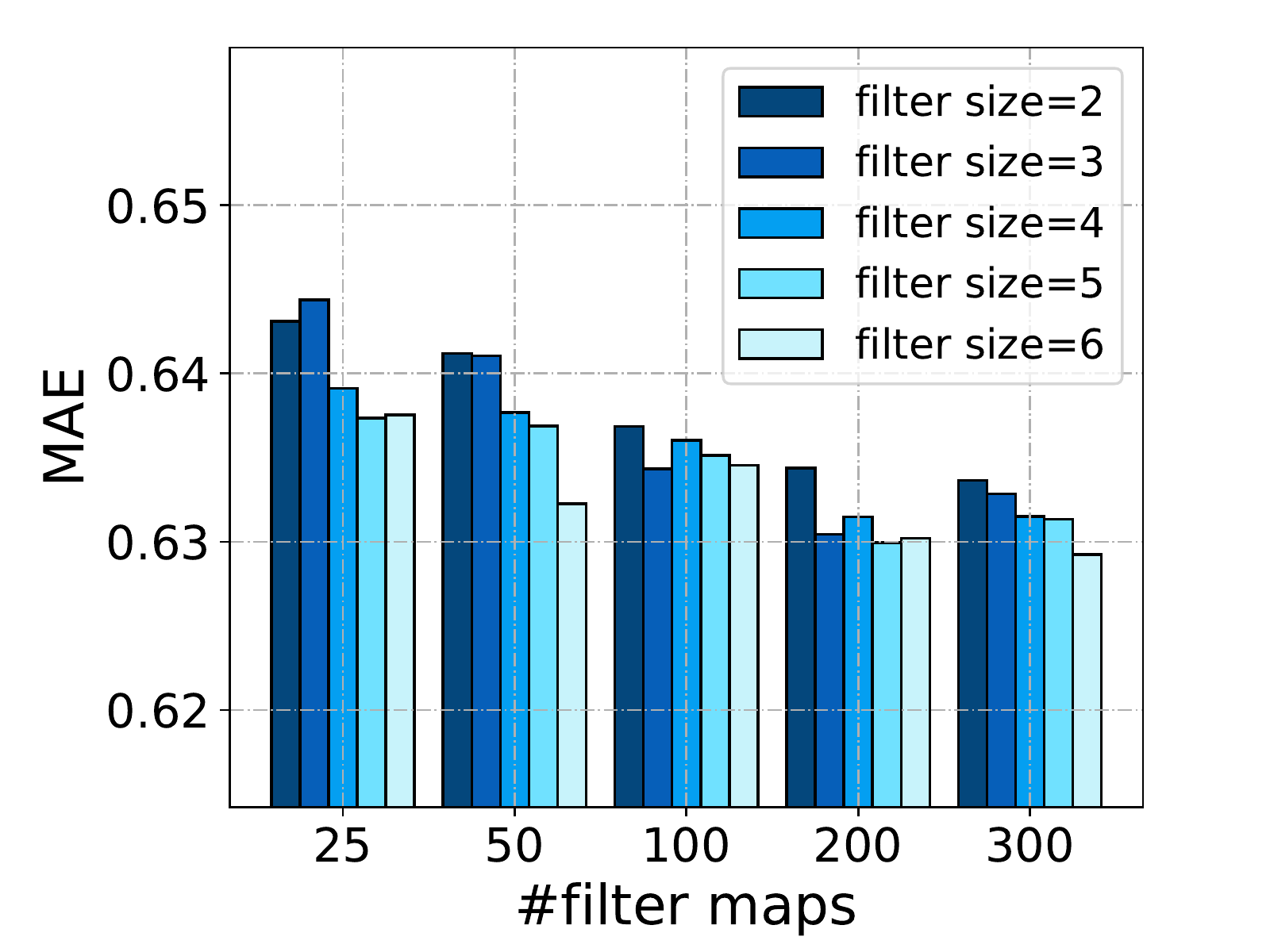}}

  \caption{MAE with respect to different combinations of filter size and number of filter maps across the datasets. }
  \label{fig:c-params}
\end{figure*}

The CNNs for extracting local semantics have two important parameters, namely filter size, and the number of filter maps. We vary the filter size amongst [2,3,4,5,6] for the first and second layer CNNs. As for the number of filter maps, we vary the hyper-parameter amongst [25,50,100,200,300] for the first layer CNN and set it equal to the number of latent factors for the second. As depicted in Fig.\ref{fig:c-params}, we only show the MAE with respect to different CNNs parameters, as RMSE presents the similar trend. We can see i) the proposed model experiences an average of 2.13\% in performance variants under different CNNs parameters, demonstrating its robustness for achieving consistent recommendation performance; ii) the proposed model generally achieves the optimal results with larger filter size and more filter numbers, showing that more parameters are required for increasing the model representativeness and expressiveness; iii) the performance on \textit{Musical Instruments} and \textit{Office Products} begins to degrade with more filter maps, possibly because the model gradually overfits the small datasets with complex CNNs parameters.

\subsubsection{Time complexity analysis}
\begin{table*}\footnotesize
\centering
\caption{\normalsize{Running time in seconds of the competitive models across the datasets. The training time is the time for training an epoch of the data.}}
\begin{tabular}{l|cc|cc|cc|cc|cc}
\toprule
\multirow{2}{*}{Models} & \multicolumn{2}{|c|}{\textit{Musical Instruments}} & \multicolumn{2}{c|}{\textit{Office Products}} & \multicolumn{2}{c|}{\textit{Grocery and Gourmet Food}} & \multicolumn{2}{c|}{\textit{Video Games}} & \multicolumn{2}{c}{\textit{Sports and Outdoors}} \\
\cline{2-11}
& train & test & train & test & train & test & train & test & train & test \\
\hline
PMF & 0.616 & 0.031 & 3.370 & 0.142 & 10.098 & 0.406 & 15.399 & 0.768 & 20.448 & 1.022\\
NeuMF & 0.812 & 0.040 & 4.060 & 0.165 & 11.719 & 0.587 & 18.729 & 0.765 & 26.743 & 1.540\\
CDL & 2.053 & 0.091 & 10.566 & 0.584 & 31.452 & 1.218 & 53.148 & 2.500 & 70.675 & 2.677 \\
ConMF & 2.425 & 0.099 & 12.858 & 0.682 & 42.649 & 1.945 & 53.318 & 2.826 & 70.118 & 3.879\\
DeepCoNN&3.660&0.194&22.484&1.122&63.477&3.182&85.095&4.910&128.966&6.534\\
D-attn&10.118&0.590&59.501&3.128&164.936&9.124&211.647&13.608&338.507&17.912\\
CARL&4.961&0.218&29.619&1.122&86.245&3.340&110.168&4.910&172.168&6.644\\
NARRE & 4.282 & 0.250 & 21.408 & 0.954 & 74.376 & 2.872 & 108.362 & 5.239 & 144.203 & 7.196\\
DAML & 4.983 & 0.193 & 25.379 & 1.212 & 87.672 & 4.525 & 114.934 & 6.484 & 171.536 & 9.489\\
HTI&5.365&0.219&26.825&1.262&77.425&3.853&117.943&5.893&151.441&7.586\\
\bottomrule
\end{tabular}
\label{tb:time}
\end{table*}
To study the time efficiency of the proposed model, we compare its running time with some representative baselines. We record the running time across the models on the same computing environment, and set the hyperparameters according to the original works. Table.\ref{tb:time} shows the training and prediction time for the deep models. The training time is the time for training one epoch of the data, and the prediction time is the time required to complete the prediction for the whole testing set. We can observe that PMF and NeuMF take the least time to finish the training and testing, as those two models do not involve text-related tasks. CDL and ConMF only incorporate simple text modeling on the item side, and they incur the least overhead among the text-based models.  D-attn needs to calculate local attentions with  kernels of various sizes, leading to the slowest model. CARL and DAML share the similar architecture, therefore, they have similar training and testing costs. NARRE is a hierarchical model, however, it does not model hierarchical text alignments. Therefore, NARRE runs faster than the proposed model in terms of training and testing. Comparing HTI with the state-of-the-art DAML, we can see the overhead differences between them are data-dependent. For example, DAML takes less time than HTI on \textit{Musical Instruments} and \textit{Video Games}. By contrast, HTI yields lower overhead than DAML on \textit{Grocery and Gourmet Food} and \textit{Sports and Outdoors}. The experiment shows that HTI is able to achieve the best recommendation accuracy without incurring noticeable overhead.
\subsection{Visualization}
%
%

\begin{figure*}
  \centering
  \subcaptionbox{1st review (5.0)}[.24\textwidth][c]{%
    \includegraphics[width=.24\textwidth]{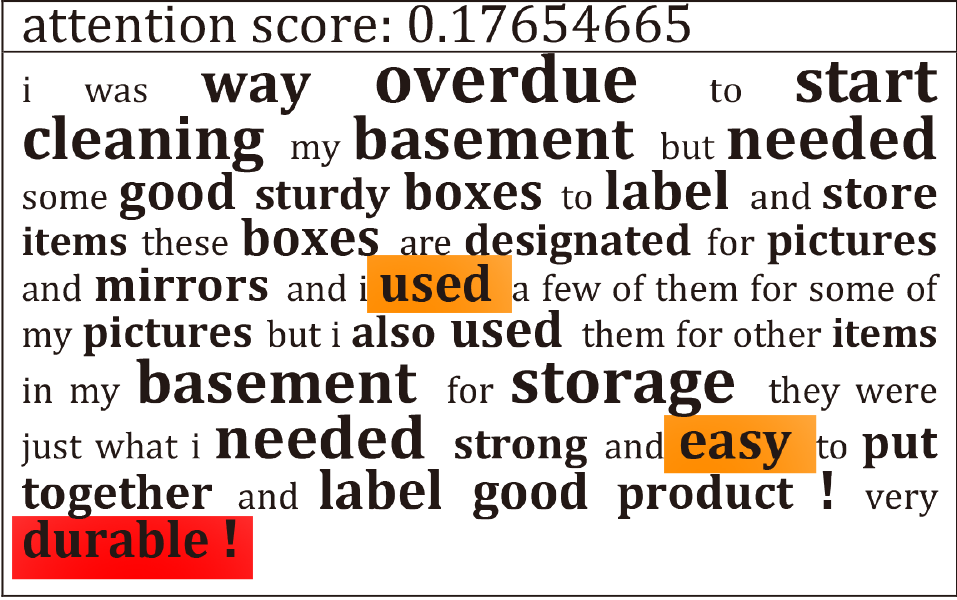}}
  \subcaptionbox{2nd review (5.0)}[.24\textwidth][c]{%
    \includegraphics[width=.24\textwidth]{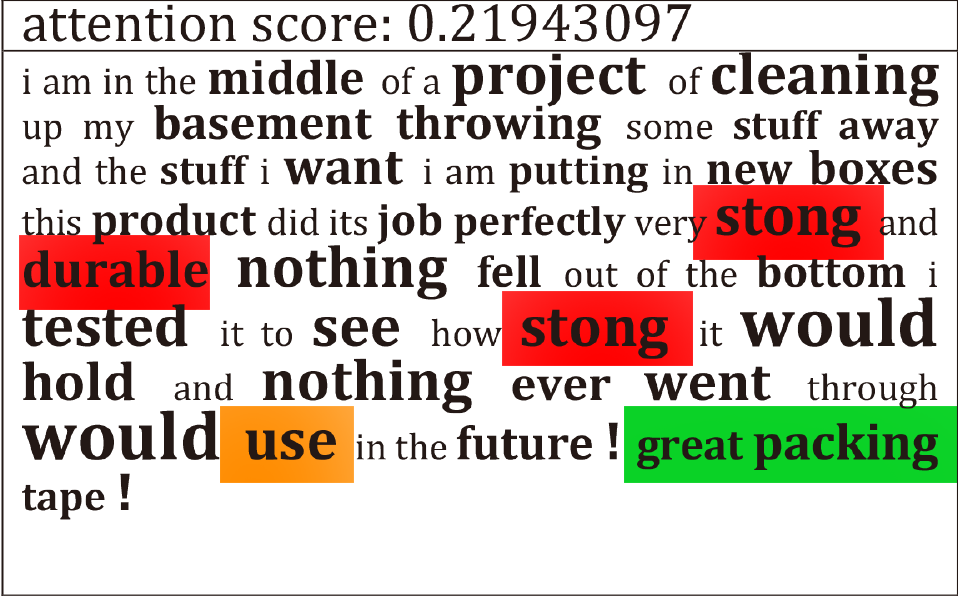}}
  \subcaptionbox{3rd review (5.0)}[.24\textwidth][c]{%
    \includegraphics[width=.24\textwidth]{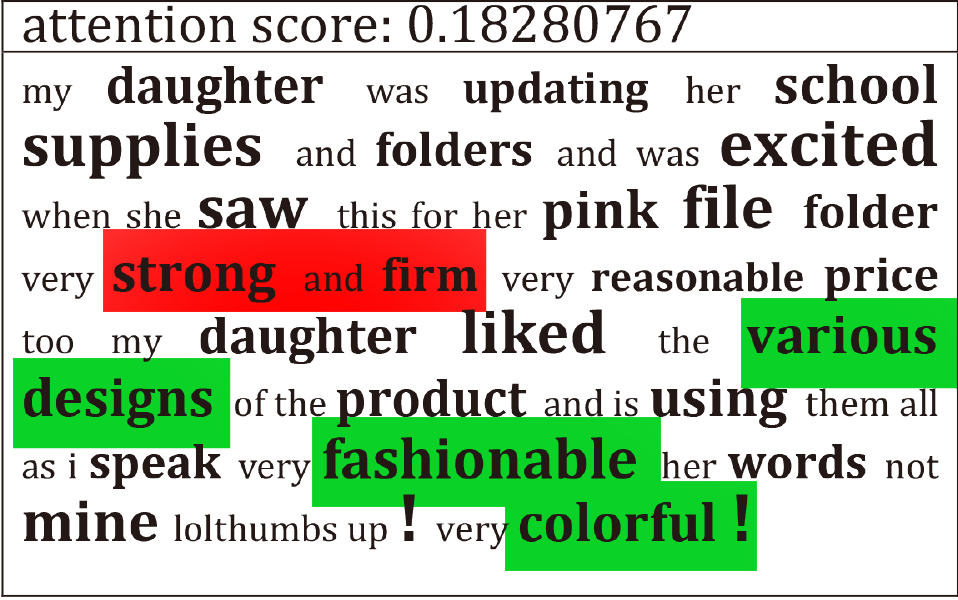}}
  \subcaptionbox{4th review (4.0)}[.24\textwidth][c]{%
    \includegraphics[width=.24\textwidth]{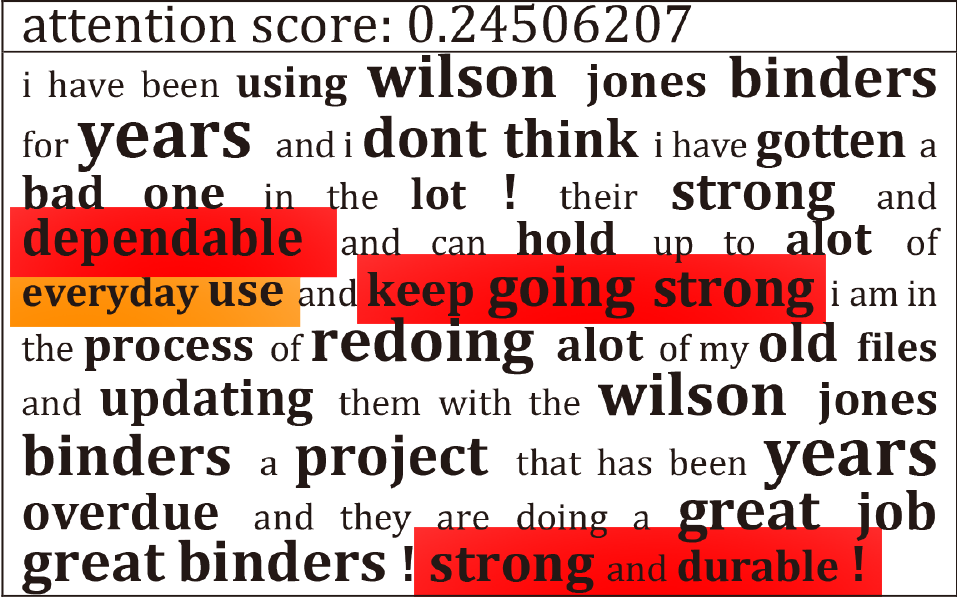}}
  \caption{Visualization of the \textbf{user reviews} given the tuple \textbf{(\#4492, \#1027, 5.)}, where the ground truth rating is 5 \textbf{(positive rating)}. The attentions over the review are shown at the top of each review, while the attentions over the words are indicated by the font size. The ground-truth ratings of those reviews are shown in the parentheses of the sub-captions. The colors are used to present the word correlations between the user and item reviews, with each color indicating a potential aspect under which the words semantically match. }
  \label{fig:1-user}
\end{figure*}

\begin{figure*}
  \centering
  \subcaptionbox{1st review (5.0)}[.24\textwidth][c]{%
    \includegraphics[width=.24\textwidth]{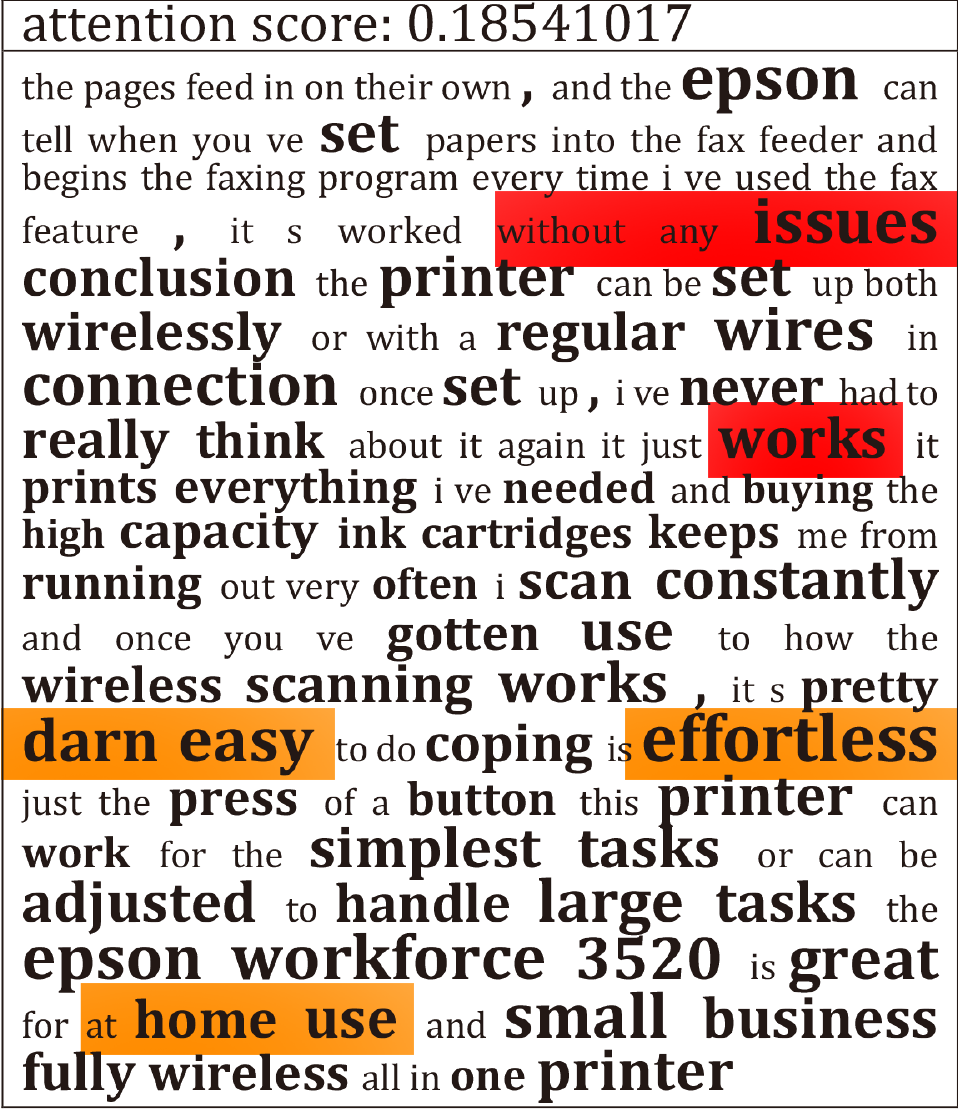}}
  \subcaptionbox{2nd review (5.0)}[.24\textwidth][c]{%
    \includegraphics[width=.24\textwidth]{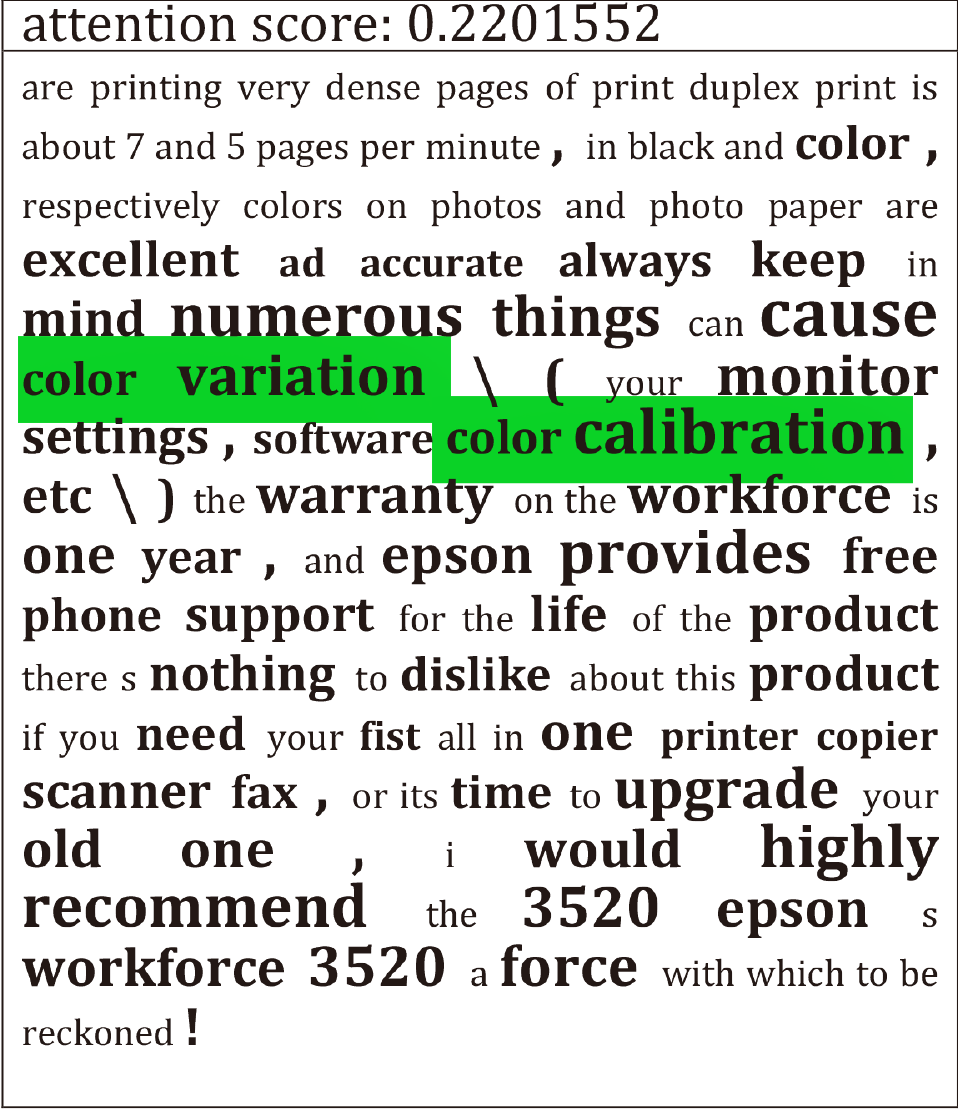}}
  \subcaptionbox{3rd review (5.0)}[.24\textwidth][c]{%
    \includegraphics[width=.24\textwidth]{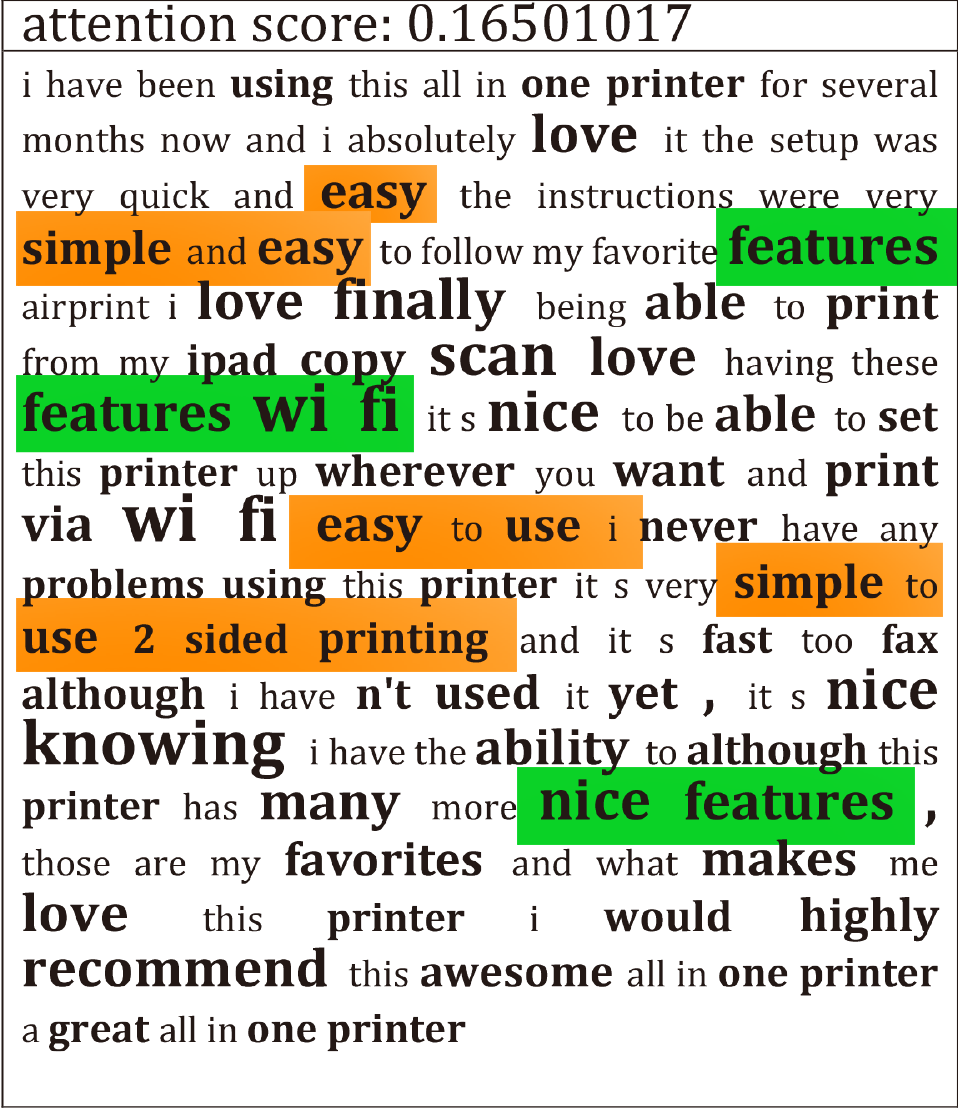}}
  \subcaptionbox{4th review (5.0)}[.24\textwidth][c]{%
    \includegraphics[width=.24\textwidth]{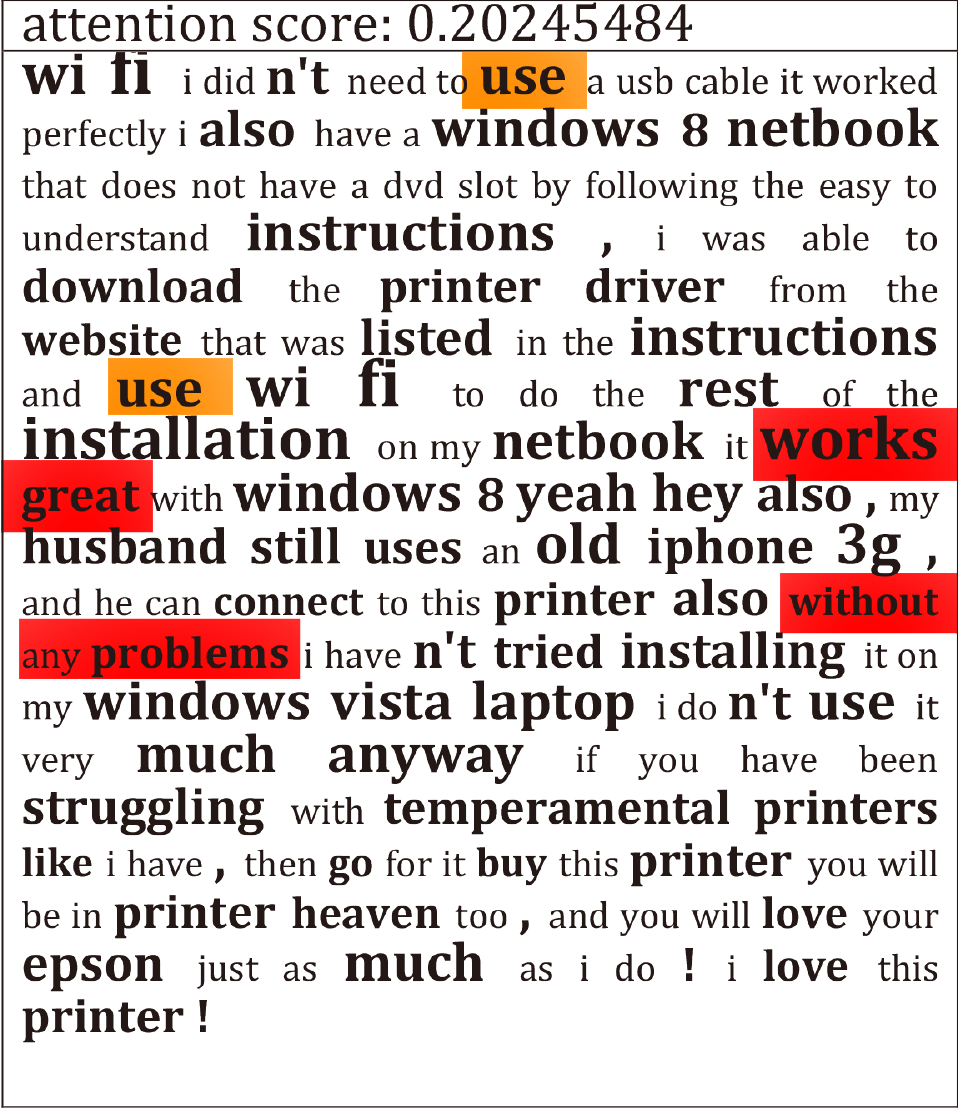}}
  \caption{Visualization of the \textbf{item reviews} given the tuple \textbf{(\#4492,\#1027, 5.)}, where the ground truth rating is 5 \textbf{(positive rating)}. The attentions over the review are shown at the top of each review, while the attentions over the words are indicated by the font size. The ground-truth ratings of those reviews are shown in the parentheses of the sub-captions. }
  \label{fig:1-item}
\end{figure*}

\begin{figure*}
  \centering
  \subcaptionbox{1st review (1.0)}[.24\textwidth][c]{%
    \includegraphics[width=.24\textwidth]{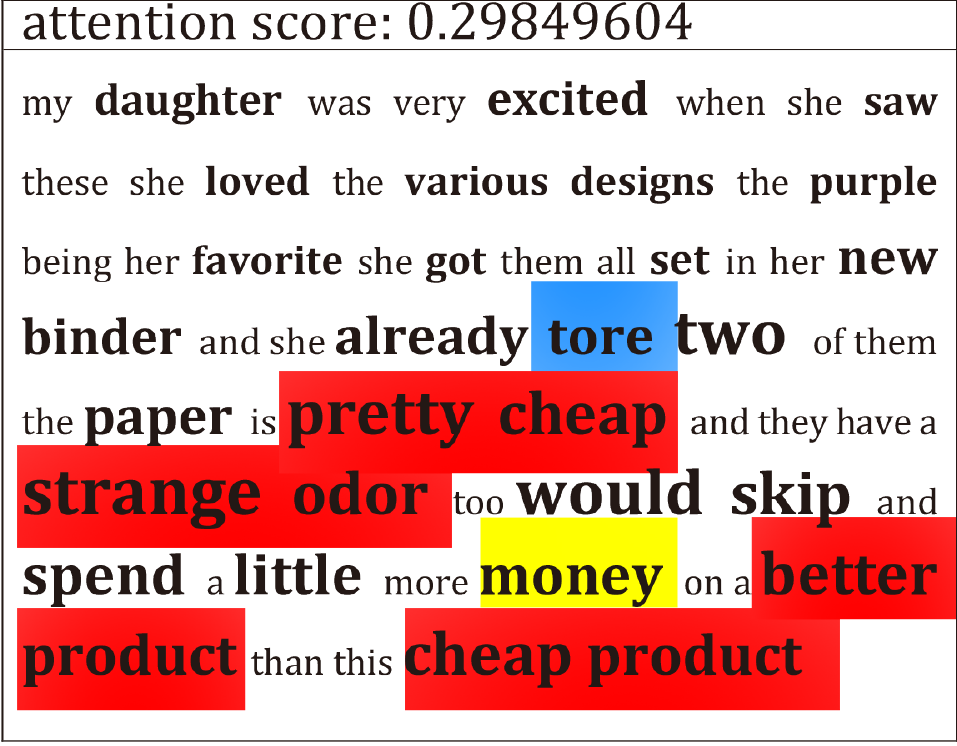}}
  \subcaptionbox{2nd review (1.0)}[.24\textwidth][c]{%
    \includegraphics[width=.24\textwidth]{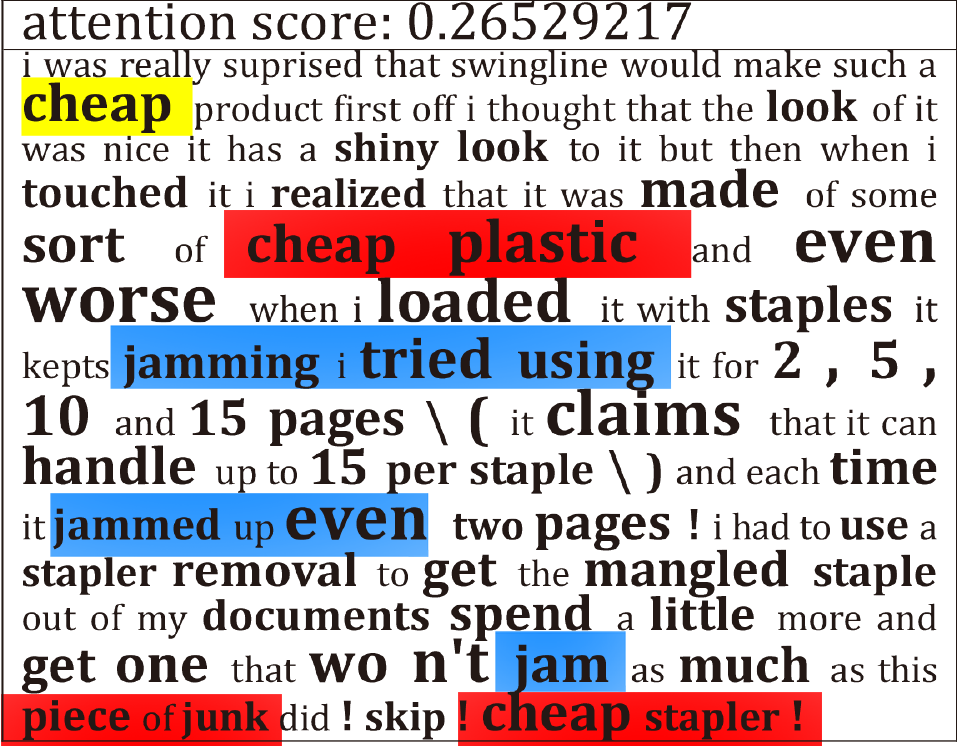}}
  \subcaptionbox{3rd review (5.0)}[.24\textwidth][c]{%
    \includegraphics[width=.24\textwidth]{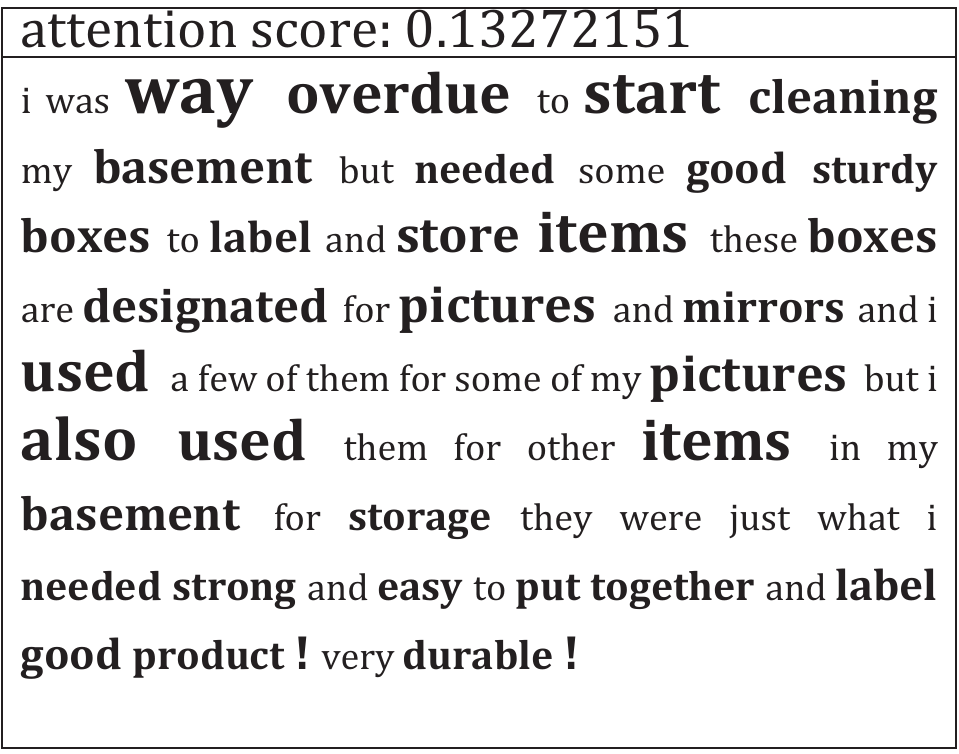}}
  \subcaptionbox{4th review (5.0)}[.24\textwidth][c]{%
    \includegraphics[width=.24\textwidth]{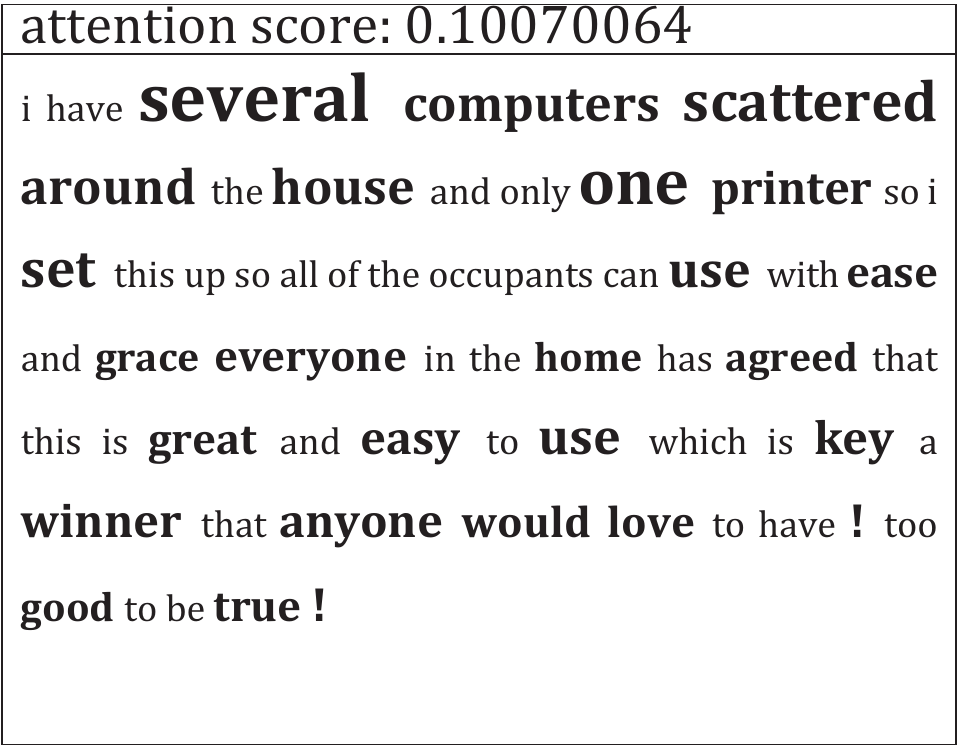}}
  \caption{Visualization of the \textbf{user reviews} given the tuple \textbf{(\#4492, \#856, 1.)}, where the ground truth rating is 1 \textbf{(negative rating)}. The attentions over the review are shown at the top of each review, while the attentions over the words are indicated by the font size. The ground-truth ratings of those reviews are shown in the parentheses of the sub-captions. Notice that the later two reviews have opposite sentiment, however, they are placed with low weights. Therefore, we focus our discussions on the former two reviews.}
  \label{fig:2-user}
\end{figure*}

\begin{figure*}
  \centering
  \subcaptionbox{1st review (2.0)}[.24\textwidth][c]{%
    \includegraphics[width=.24\textwidth]{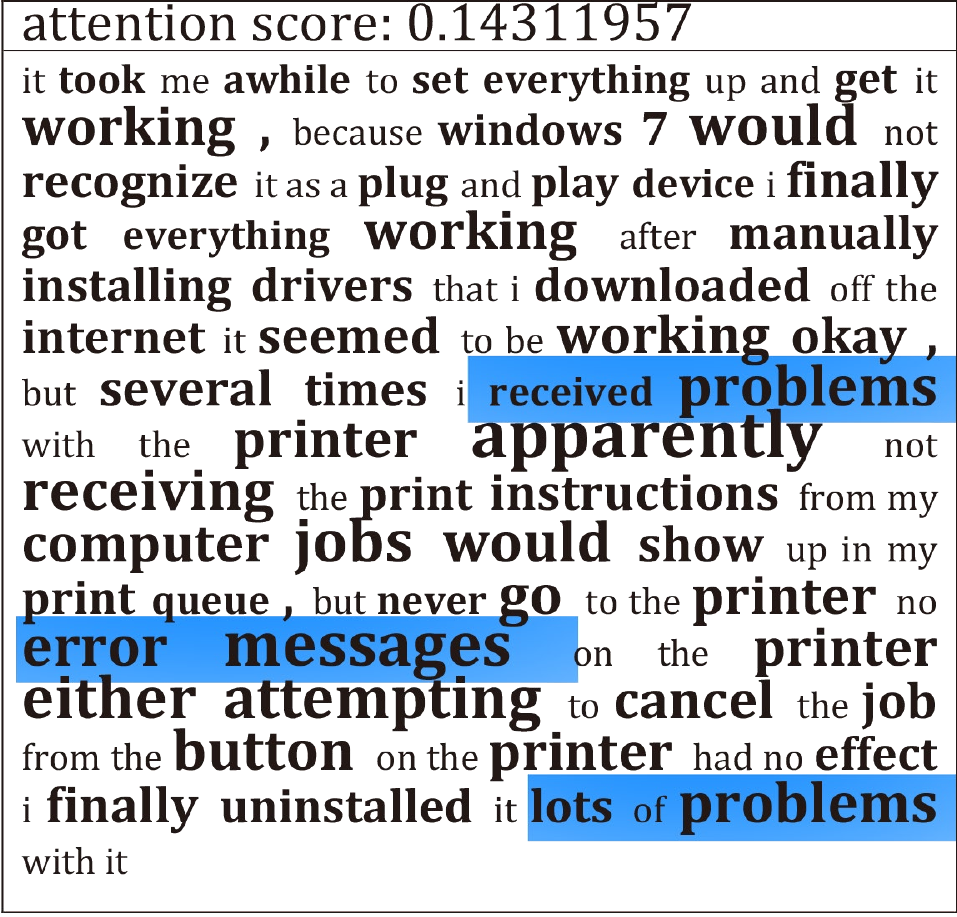}}
  \subcaptionbox{2nd review (1.0)}[.24\textwidth][c]{%
    \includegraphics[width=.24\textwidth]{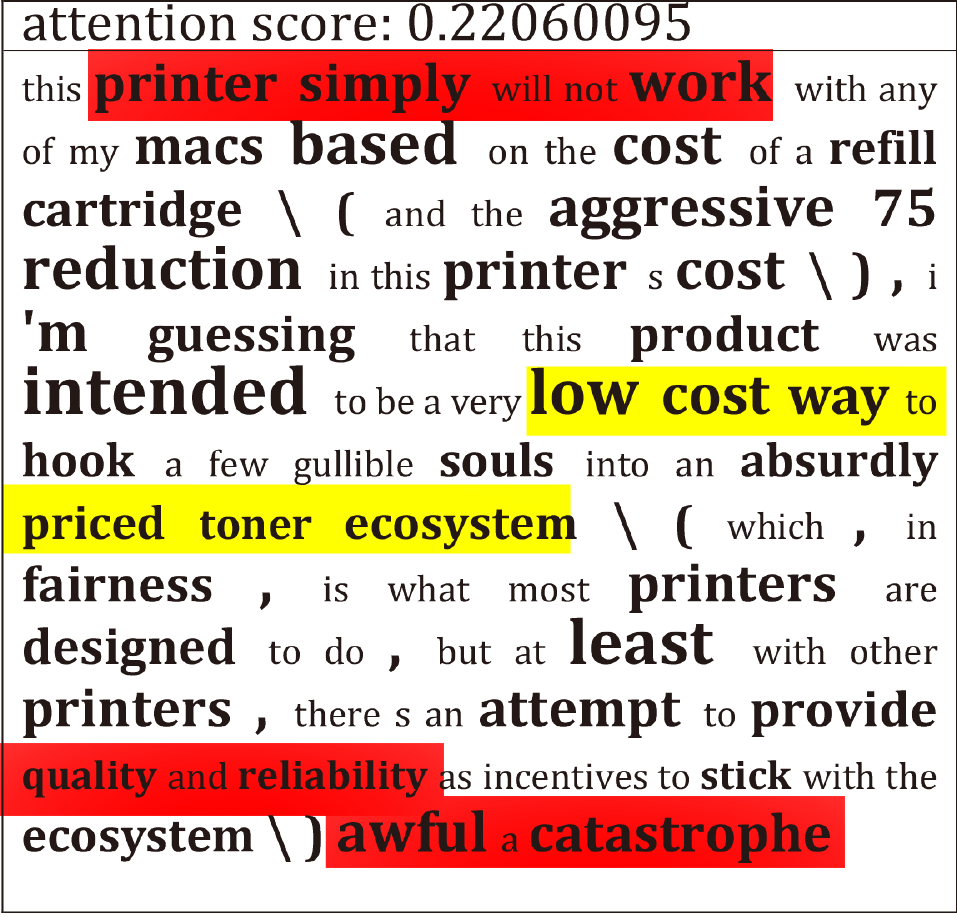}}
  \subcaptionbox{3rd review (1.0)}[.24\textwidth][c]{%
    \includegraphics[width=.24\textwidth]{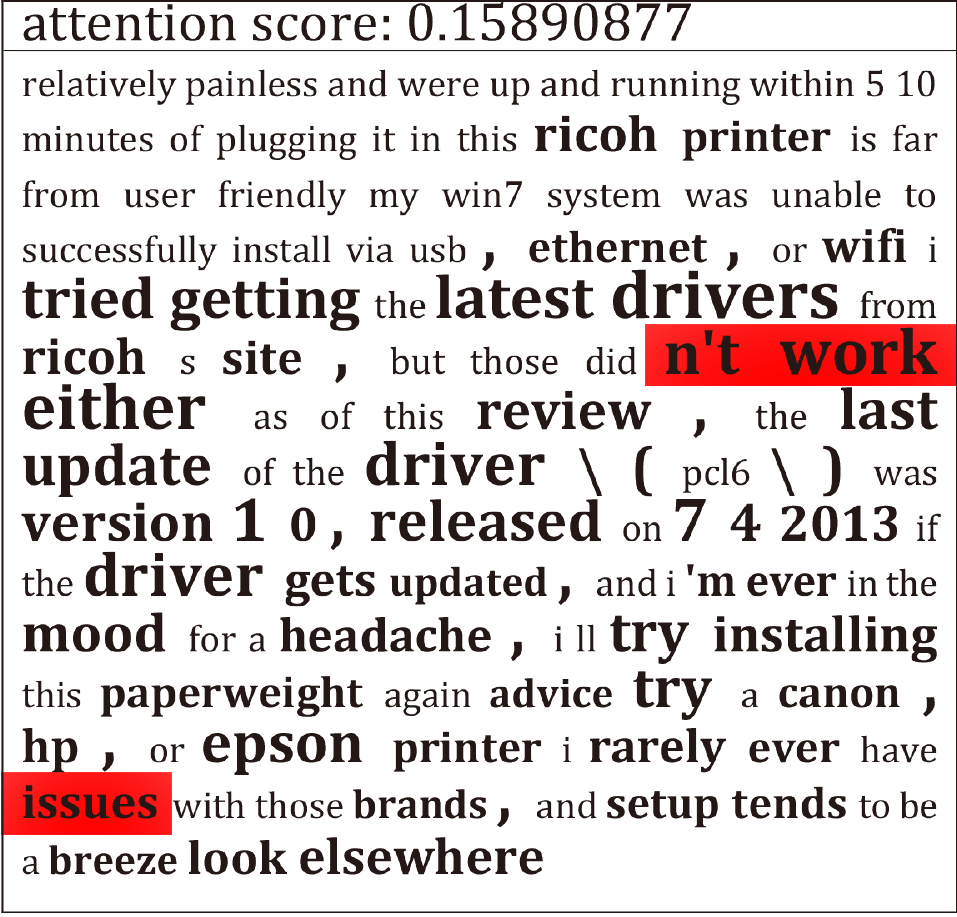}}
  \subcaptionbox{4th review (1.0)}[.24\textwidth][c]{%
    \includegraphics[width=.24\textwidth]{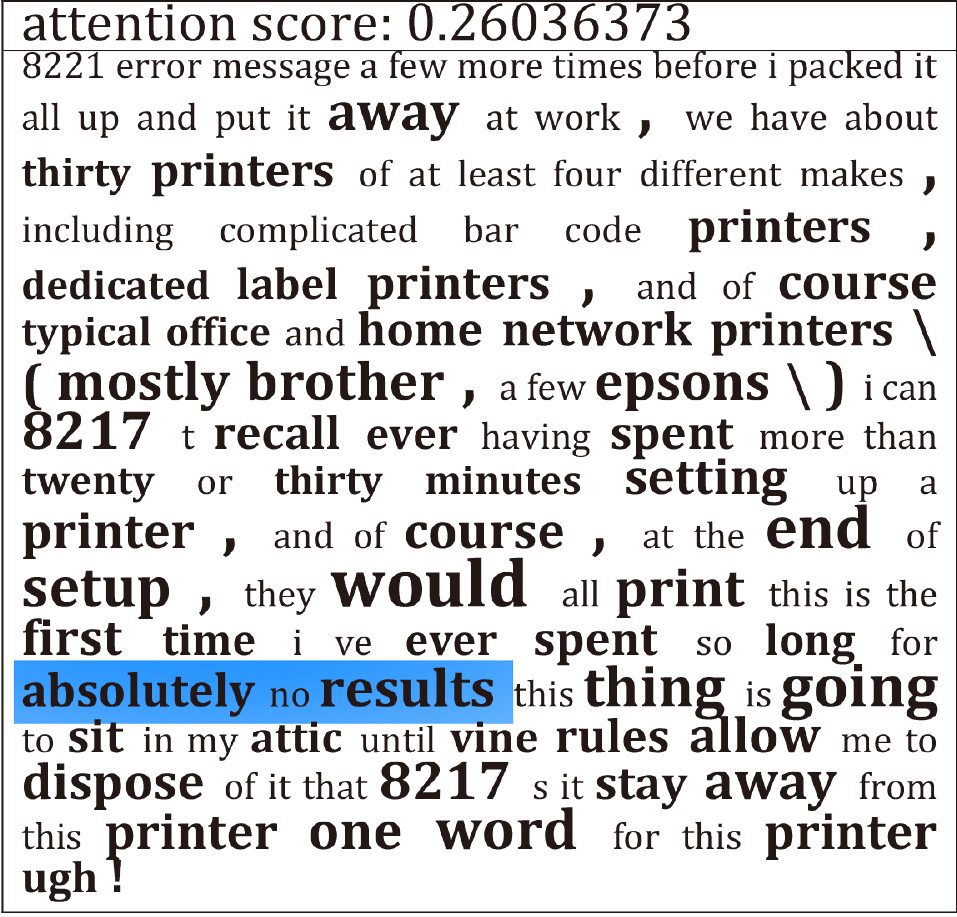}}
  \caption{Visualization of the \textbf{item reviews} given the tuple \textbf{(\#4492, \#856, 1.)}, where the ground truth rating is 1 \textbf{(negative rating)}. The attentions over the review are shown at the top of each review, while the attentions over the words are indicated by the font size. The ground-truth ratings of those reviews are shown in the parentheses of the sub-captions.}
  \label{fig:2-item}
\end{figure*}

\begin{table}
\centering
\caption{The correspondence between the colors in Fig.\ref{fig:1-user}-\ref{fig:2-item} and the potential aspect under which the words are correlated.}
\begin{tabular}{p{1.5cm}c}
\toprule
color & potential aspect \\
\hline
\begin{minipage}[b]{0.1\textwidth}\raisebox{-.35\height}{\includegraphics[width=1.5cm]{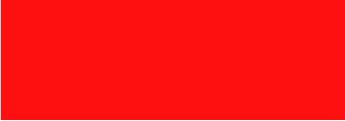}} \end{minipage} & quality\\
\begin{minipage}[b]{0.1\textwidth}\raisebox{-.35\height}{\includegraphics[width=1.5cm]{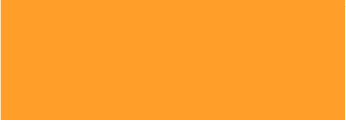}} \end{minipage} & usability\\
\begin{minipage}[b]{0.1\textwidth}\raisebox{-.35\height}{\includegraphics[width=1.5cm]{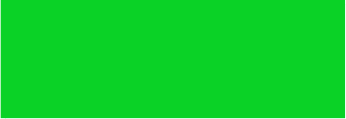}} \end{minipage} & feature\\
\begin{minipage}[b]{0.1\textwidth}\raisebox{-.35\height}{\includegraphics[width=1.5cm]{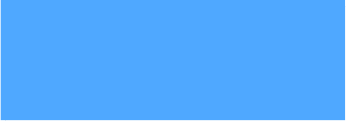}} \end{minipage} & functionality\\
\begin{minipage}[b]{0.1\textwidth}\raisebox{-.35\height}{\includegraphics[width=1.5cm]{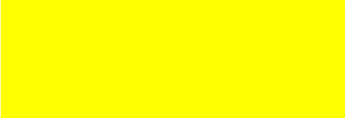}} \end{minipage} & price\\
\bottomrule
\end{tabular}
\label{tb:aspect}
\end{table}

%
%


In this subsection, to investigate whether HTI can capture text correlations at different levels of hierarchies, we visualize the attention scores both in word-level and review-level. We randomly sample two rating records having the same user, but with different rating scores. A rating record is denoted as $(u_i, v_j, r_{ij})$ such that the user $u_i$ rates the item $v_j$ with a score $r_{ij}$. We visualize the top-4 reviews having the largest attention scores. The attention scores are visualized by the fontsize of the words in the reviews, with large fontsize representing high attentions and small fontsize denoting low values.

Fig.\ref{fig:1-user} and Fig.\ref{fig:1-item} show the visualizations of user and item reviews of the first tuple (\#4492,\#1027, 5), respectively. Fig.\ref{fig:2-user} and Fig.\ref{fig:2-item} show the visualizations of the user and item reviews of the second tuple (\#4492, \#856, 1), respectively. The attention scores are list at the top of each review, while the ground-truth ratings (labels) are shown in the parentheses of the review captions.

For the word-level visualization, we can see that informative words are placed with high attention scores across the reviews. For example, in Fig.\ref{fig:1-user}, the words ``durable", ``great packing", ``fashionable", ``dependable" are assigned with large attention scores. Those words are informative, as they have strong indications of the user's preference. In Fig.\ref{fig:1-item}, the words ``easy", ``nice features", ``work great" receive high attentions. Those words reflect different characteristics of the item.

As for the review-level visualization, given the positive tuple (\#4492,\#1027, 5), it is clear to see that the proposed model can place higher attention scores to positive reviews for rating prediction, as shown in Fig.\ref{fig:1-user} and Fig.\ref{fig:1-item}. Similarly, given the negative tuple (\#4492, \#856, 1), we are able to locate the negative reviews for recommendation, as presented in Fig.\ref{fig:2-user} and Fig.\ref{fig:2-item}. An exception can be found in the user reviews of the negative tuple, where the first two reviews show negative sentiment, while last two reviews present positive sentiment. However, HTI attributes high attention scores to the first two reviews and low scores to the others, demonstrating that HTI can capture informative textual features at the review level given a specific user-item pair.

It is also interesting to find the word correlations between the user and item reviews. In Fig.\ref{fig:1-user}-\ref{fig:2-item}, we use color to highlight word correlations, with each color indicating a potential aspect under which the word are correlated. The mappings between the colors and the potential aspects are shown in Table.\ref{tb:aspect}. Specifically, for the positive tuple (\#4492,\#1027, 5), red words such as ``durable" and ``problem" imply the semantic interrelations between the user and item with respect to the potential quality aspect, while orange words such as ``easy" and ``use" depict semantic matching under the potential usability aspect. As for the negative tuple (\#4492, \#856, 1), we can find semantic correlations between the user and item in the functionality aspect, as indicated by the blue words such as ``jamming", ``tore" and ``error message".

In this case study, we can see that HTI can highlight words with strong sentimental indications at the word level. At the review level, HTI can dynamically assign different attention scores to the reviews given different items. In addition, it is able to capture semantic correlations at review level for prediction. To conclude, the proposed model can exploit information textual features at different levels of hierarchies, and capture the mutual semantic correlations between specific user-item pairs to derive accurate recommendations.

\section{Conclusion and Future Work}
In this article, we propose a novel hierarchical text interaction model, named HTI, that exploits textual information from reviews for rating prediction. The essence of HTI is to model semantic correlations at different level of granularities. Two core components of HTI are word-level attention module and review-level interaction module. The former captures important words for comprising a review, while the later models text interaction and identifies the informative reviews. The experimental results on five real-world datasets show that HTI achieves significantly better recommendation accuracies than the state-of-the-art models. In addition, we validate the effectiveness of the hierarchical architecture. The validation is completed by comparing HTI with its variants that exclude the attention at word level and interaction at review level. To provide a better understanding of HTI's efficacy in capturing semantic correlations at different granularities, we visualize attentions at both word and review levels for randomly sampled user-item pairs.

In this article, we mainly focus on hierarchical text interaction for improving recommendation performance. However, in the proposed model, we linearly combine text representation with latent factors for rating prediction, ignoring personalized text usefulness among the users, as not all users equally depend on the reviews for the recommendation. In addition, the proposed HTI model only involves single data modality for the recommendation task. As part of future work, we plan to investigate the aforementioned factors for rating prediction, such as the modeling of personalized text usefulness, interaction across heterogeneous data modalities \cite{TCYB2019_cite_leizhu}, etc.

\section{Acknowledgement}
This research was supported in part by the National Natural Science Foundation of China under Grant 61802427.

\bibliography{bib}

\end{document}